\documentstyle[12pt]{article}
\newcommand{\fslash}[1]{\mbox{$\!\not\!#1$}}

\newcommand{\be}{\begin{equation}}
\newcommand{\ee}{\end{equation}}

\newcommand{\bold}[1]{\mbox{\boldmath ${#1}$}}

\begin{document}

\baselineskip 4 ex

\title{Phase transition from nuclear matter to color superconducting quark matter}
       
\author{W. Bentz \thanks{Correspondence to: W. Bentz, E-mail: bentz@keyaki.cc.u-tokai.ac.jp}, 
       T. Horikawa \\
       Department of Physics, School of Science, Tokai University \\
       Hiratsuka-shi, Kanagawa 259-1292, Japan \\
{ } \\
       N. Ishii \\
       The Institute of Physical and Chemical Research (RIKEN) \\
       Hirosawa, Wako-shi, Saitama 351-0198, Japan \\      
{ } \\
       A.W. Thomas \\
       Special Research Centre for the Subatomic Structure of Matter \\
       and \\
       Department of Physics and Mathematical Physics,\\
       The University of Adelaide  \\
       Adelaide, SA 5005, Australia}

\date{ }
\maketitle
\newpage
\begin{abstract}
We construct the nuclear and quark matter equations of state at zero temperature
in an effective quark theory (the Nambu-Jona-Lasinio model), and discuss the phase transition 
between them. The nuclear matter equation of state is based on the quark-diquark
description of the single nucleon, while the quark matter equation of state includes the effects
of scalar diquark condensation (color superconductivity). The effect of diquark condensation
on the phase transition is discussed in detail.  
\\

\noindent

{\small 
}
\end{abstract}

\newpage

\section{Introduction}
\setcounter{equation}{0}
The behavior of matter at high baryon density is of great interest in
connection with neutron stars, the possible
existence of quark stars \cite{STAR}, and heavy ion collisions \cite{HI}. At normal densities,
hadronic matter consists of three-quark (nucleon-like \cite{Guichon:1995ue}) clusters, and the forces between them
lead to the familiar saturation properties of nuclear matter (NM). 
At high densities one expects a phase transition to quark matter (QM) \cite{PT}, but the
dynamics of this transition, the transition densities, and the equation of state (EOS) 
in the high density regime are largely unknown. In many recent investigations it has been found
that the QM state at high densities differs from a non-interacting quark gas. In particular, there is an
instability with respect to diquark condensation, leading to a color superconducting 
state \cite{CS1}-\cite{CS3}. On the other hand, diquark correlations are known to play an 
important role in explaining
the properties of single nucleons, for example the flavor dependence of the structure
functions \cite{CT,MIN}. At high baryon densities one therefore expects a phase transition from
quark-diquark clusters to a diquark condensed phase. 

Since it is still difficult to obtain unambiguous results for finite density
matter directly from QCD, effective quark models provide powerful tools to
investigate the behavior of matter as the baryon density increases. Naturally, these models 
should account for the properties of the free nucleon in terms of
constituent quarks, as well as for the  saturation properties of nuclear matter at normal densities.
They should also allow for the possibility
of a high density QM phase, where chiral symmetry is restored and color symmetry 
spontaneously broken.
In these respects, the Nambu-Jona-Lasinio (NJL) model \cite{NJL,NJL1} is an attractive 
candidate. First, the simplicity of the model allows a direct solution of the relativistic
Faddeev equation \cite{ISH}, leading to a covariant and successful description of the 
single nucleon \cite{FAD1}-\cite{MIN1}. 
Second, it has been shown recently \cite{BT} that this successful
quark-diquark description of the single nucleon can be extended to describe the EOS
of NM. The essential ingredient needed to obtain a saturating EOS was to simulate the
effect of confinement by introducing
an infrared cut-off (in addition to the usual ultraviolet one) in order to
avoid unphysical thresholds corresponding to the decay of hadrons into free quarks \cite{IR}. Third,
models of the NJL-type are able to describe a high density QM 
state where chiral symmetry is restored and color symmetry
 is broken \cite{CS1}. It is, of course, 
desirable to combine these three aspects and describe nucleons, NM and superconducting
QM in one consistent framework. This is the purpose of our present work. 

In this paper we will investigate the role of diquark
condensation for the phase transition from NM to QM in the NJL model. It is clear that
diquark condensation will soften the EOS of QM and favor the phase transition. The
transition densities are therefore expected to decrease with increasing pairing
strength. We will investigate this relation, and 
discuss a possible scenario for the phase transition. In the course of the calculation
we will face the problem of whether or not those parameters of the model which have been 
determined exclusively
by the properties of single nucleons and NM, like the value of the infrared cut-off, the strength of the 
$q \overline{q}$ interaction in the vector meson channel, and the strength of the $qq$ 
interaction in the scalar diquark channel,
should be taken over to QM. In order to discuss these points, in particular the strength of the 
$q \overline{q}$ interaction in the vector meson channel, we will also consider the $\omega$ meson mass
in the medium. Another purpose of this work is to present a consistent formulation
which accounts for the three aspects discussed above -  i.e., the single
nucleon, normal NM and superconducting QM.     
 
In sect. 2 we will first explain the model in the framework of the Nambu-Gorkov 
formalism \cite{NG1,NG2}, and then use path integral methods to describe the 
single nucleon as well as the EOS for NM and QM.
In sect. 3 we will present present numerical results, and discuss the
role of diquark condensation in 
the phase transition. Finally, we summarize our results in sect. 4.
         
\section{NJL model in the Nambu-Gorkov formalism}

The Lagrangian density of the flavor SU(2) NJL model has the general
form ${\cal L}=\overline{\psi}\left(i\fslash{\partial}-m\right)\psi +
\sum_{\alpha} G_{\alpha}\left(\overline{\psi} \Gamma_{\alpha}\psi\right)^2$,
where $m$ is the current quark mass, and the chirally symmetric 4-fermi interaction is characterized by
the matrices $\Gamma_{\alpha}$ in Dirac, flavor and color space, and the corresponding coupling constants 
$G_{\alpha}$. Hadronization techniques to describe the single nucleon and NM, as well as the description of 
superconducting QM, are simplified very much if one makes use of the Nambu-Gorkov formalism \cite{NG1,NG2}.   
In this method, the Lagragian density is re-written identically in terms of the fields
\footnote{Since there is freedom under unitary transformations to define the field $\Psi$, 
the presence of the matrices 
$C$ and $\tau_2$ in the lower components is a matter of definition. We include them here
because they lead to simplifications in some of the following formulae.}
\begin{equation}
\Psi = \frac{1}{\sqrt{2}} \left(
\begin{array}{cc}
\psi \\ C \tau_2 \overline{\psi}^T
\end{array} \right), \,\,\,\,\,\,\,\,\,\,\,\,\,\,\,
\overline{\Psi} = \frac{1}{\sqrt{2}} \left(
\overline{\psi},\,\, - \psi^T \tau_2 C^{-1} 
\right), \label{fields}
\end{equation}
where $C=i \gamma_0 \,\gamma_2$, as follows:
\begin{equation}
{\cal L}=\overline{\Psi}\left(i\fslash{\partial}-m\right)\Psi +
\sum_{\alpha\,M} G_{\alpha}\left(\overline{\Psi} \Gamma_{\alpha} \sigma_{M} \Psi\right)^2. \label{nglag}
\end{equation}
We denote the matrices in the Nambu-Gorkov space by $\sigma_M\equiv (\sigma_0, \sigma_3)$ for the 
interactions in mesonic ($q{\overline q}$) channels,
and by $\sigma_D\equiv (\sigma_1, \sigma_2)$ for the diquark ($qq$) channels. Here $\sigma_0$ is the
$2 \times 2$ unit matrix, and 
$\sigma_1, \sigma_2$, $\sigma_3$ are the usual Pauli matrices. It is easy to
verify that the only non-zero terms in (\ref{nglag}) are those which satisfy 
\begin{equation}
C \,\tau_2\, \sigma_2\, \Lambda_{\alpha}^T\, C^{-1}\, \sigma_2 \,\tau_2 = \Lambda_{\alpha}, 
\label{cond}
\end{equation}
where $\Lambda_{\alpha} \equiv \Gamma_{\alpha} \sigma_M$. Therefore, for a given $\Gamma_{\alpha}$, either
$M=0$ or $M=3$ contributes in (\ref{nglag}).

Any Lagragian density of the form (\ref{nglag}) can further be re-written into a Fierz symmetric form
\cite{ISH}, where the 
interactions in the diquark channels, characterized by $\sigma_{D}$, also appear.
For these terms to be non-zero, the same condition (\ref{cond}) holds with 
$\Lambda_{\alpha}\equiv \Gamma_{\alpha}\sigma_D$. 
The coupling constants appearing in the Fierz
symmetric Lagrangian density for the various meson and diquark channels are linear 
combinations of the $G_{\alpha}$'s
appearing in the Lagrangian density (\ref{nglag}). For the purposes of this work, 
we will need only the following terms:
\begin{eqnarray}
{\cal L}_I &=& G_{\pi} \left(\left(\overline{\Psi} \Psi \right)^2 + 
\left(\overline{\Psi} i\,\gamma_5 {\bold \tau} \sigma_3 \Psi \right)^2 \right)  \nonumber \\
&-& G_{\omega} \left(\overline{\Psi} \gamma^{\mu} \sigma_3 \Psi \right)^2 
+G_s \left(\overline{\Psi} i\,\gamma_5 \sigma_D \beta_a \Psi \right)^2, 
\label{inter}
\end{eqnarray}   
which are the interactions in the $(J^P, T) = (0^+,0), (0^-, 1)$ and $(1^-,0)$ color singlet 
meson channels, and the $(0^+,0)$ color $\overline{3}$ diquark (scalar diquark) channel.
The color $\overline{3}$ matrices are defined in terms of the antisymmetric Gell-Mann matrices by 
$\beta_a = \sqrt{3/2}\, \lambda_A$, where $a=1,2,3$ corresponds to $A=2,5,7$. 
A sum over $D=1,2$ and $a=1,2,3$ is implied in the last term of Eq.(\ref{inter}).

The 3 coupling constants in Eq.(\ref{inter}) will be treated as free parameters. The values of the ratios
\begin{equation}
r_s \equiv \frac{G_s}{G_{\pi}}, \,\,\,\,\,\,\,\,\,\,\,\,
r_{\omega} \equiv \frac{G_{\omega}}{G_{\pi}} \label{rat}
\end{equation}
reflect the form of the original interaction Lagrangian density (\ref{nglag}). 

The 4-fermi interactions, $G_{\alpha} \left(\overline{\Psi} \Lambda_{\alpha} \Psi\right)^2$, 
in Eq.(\ref{inter}) 
can be eliminated in favor of bosonic auxiliary fields $b_{\alpha}$ as usual 
\footnote{This procedure can also be
formulated in the path integral formalism, as will be done in sect. 3 for the nucleon
auxiliary field - see Eq.(\ref{add}).}
by subtracting terms of the form 
${\displaystyle \left(2 G_{\alpha} \overline{\Psi} \Lambda_{\alpha} \Psi + b_{\alpha} \right)^2
/4G_{\alpha}}$,
which are formally equal to zero if we impose the constraints (equations of motion) 
$\partial {\cal L}/\partial b_{\alpha}=0$. 
In this way we obtain the partially bosonized Lagranian density in the form:
\begin{eqnarray}
{\cal L}&=&\overline{\Psi}\left(i\fslash{\partial}-m -\Sigma -i \gamma_5 {\bold \tau}\cdot {\bold \pi} \sigma_3 
- \fslash{\omega} \sigma_3 -i \gamma_5 \sigma_D \Delta_{Da} \beta_a \right)\Psi \nonumber \\
&-& \frac{1}{4 G_{\pi}} \left(\Sigma^2+{\bold \pi}^2\right) + \frac{1}{4 G_{\omega}} \omega_{\mu}^2
- \frac{1}{4 G_{s}} \Delta_{Da}^2, \label{lag}
\end{eqnarray}
where the real bosonic auxiliary fields are defined by
\begin{eqnarray}
\Sigma = -2 G_{\pi} \overline{\Psi} \Psi ; \,\,\,\,\,\,\,\,\,\,\,\,\,\,\,
{\bold \pi} = -2 G_{\pi} \overline{\Psi}i\,\gamma_5 {\bold \tau} \sigma_3 \Psi , \label{sp} \\
\omega^{\mu}= 2 G_{\omega} \overline{\Psi}\,\gamma^{\mu} \sigma_3 \Psi ; \,\,\,\,\,\,\,\,\,\,\,\,\,\,\,
\Delta_{Da} = -2 G_{s} \overline{\Psi}i\,\gamma_5 \beta_a \sigma_D \Psi . \label{od}
\end{eqnarray}
In the two terms of (\ref{lag}) involving the diquark fields, a sum over 
$D=1,2$ and $a=1,2,3$ is implied.
  
\section{Nucleons, nuclear matter, and color superconducting quark matter}
In this section we will use the Lagrangian density (\ref{lag}) to construct the single nucleon as a
quark-scalar diquark state. We then use it to obtain the EOS of both NM and QM in the Hartree
approximation. 
\subsection{Nucleons and nuclear matter}
The EOS of NM, based on the quark-diquark description of the single nucleon, has been derived in
Ref.\cite{BT} by using a hybrid model to evaluate the expectation value 
of the quark
Hamiltonian in the NM ground state, similar in spirit to the model of Guichon and collaborators
\cite{Guichon:1995ue,GUI}.
Since it is one of the aims of this paper to present a consistent formulation of the single nucleon, 
NM and QM in one framework, we will follow the path integral formalism based on the quark 
Lagrangian density (\ref{lag}).

The general method to obtain the NM EOS is as follows: One first uses hadronization methods 
\cite{HADR,FAD2} to
derive an effective hadronic Lagrangian density of the $\sigma$-model type from the quark 
Lagrangian density, and introduces a chemical
potential for the nucleons. Then one can use any of the familiar approximation schemes for 
relativistic many-nucleon systems \cite{SW,TA}.
In this paper we will limit ourselves to the Hartree approximation. For this purpose it is sufficient to derive the nucleon 
kinetic energy term in the presence of the mean meson fields, 
the c-number terms associated with the mean fields, and the ``trace-log term'' which emerges from the integration
over the quark fields. Therefore, although the method
outlined below can be used to obtain more complete expressions for the effective hadronic Lagrangian
density, we will concentrate here on the nucleon kinetic energy term in the presence of mean scalar
and vector fields.

We introduce the (color singlet) nucleon fields 
and sources by adding the following terms to the Lagrangian density (\ref{lag}):
\begin{equation}
\left[\overline{{\cal N}} + \frac{1}{\sqrt{3}} \overline{\Psi}_a \left(
\begin{array}{cc}
\Delta^*_a & 0 \\       
0 & \Delta_a 
\end{array} \right) \right] \Phi +
\overline{\Phi}  
 \left[{\cal N} + \frac{1}{\sqrt{3}} \left(
\begin{array}{cc}
\Delta_a & 0 \\       
0 & \Delta^*_a   
\end{array} \right) \Psi_a \right]. \label{add}
\end{equation}
Here the complex diquark fields are defined by $\Delta_a = \Delta_{1a} -i \Delta_{2a}$ and
$\Delta_a^* = \Delta_{1a} +i \Delta_{2a}$, and the Nambu-Gorkov fields, ${\cal N}$ 
and $\Phi$, are expressed in terms of the nucleon field, $N$, and the nucleon source, $\phi$,
in a similar manner to (\ref{fields}) for the quark fields.   
In the generating functional, the integrations over $\Phi$ and $\overline{\Phi}$ give 
functional $\delta$-functions which show that
the nucleon auxiliary fields must satisfy the constraints $N=-\psi_a \Delta_a/\sqrt{3}$ and
$\overline{N}=-\overline{\psi}_a \Delta^*_a/\sqrt{3}$. 
The integrations over the nucleon fields then give unity --
 i.e., the terms (\ref{add}) 
do not change the generating functional of the theory. 

After integration over the quark fields (see Appendix A) we obtain the effective action
\footnote{In the action, the symbol Tr includes also a functional trace. In order to avoid
too many symbols, fields (propagators) will be frequently 
considered as vectors (matrices) in function space, besides their structures in color, flavor, and Dirac space.
The notation ${\cal S}_{\rm eff}$ will be used generically for the effective action, 
irrespective of the space of fields where it operates.} 
\begin{eqnarray}
{\cal S}_{\rm eff} &=& -\frac{i}{2} {\rm Tr\,\,ln}\, S^{-1}  
+ \int {\rm d}^4x \left( - \frac{\Delta_a^* \Delta_a}{4 G_s}
- \frac{\left(\Sigma^2+{\bold \pi}^2\right)}{4 G_{\pi}} 
+ \frac{\omega_{\mu}^2}{4 G_{\omega}}\right)  \label{leff1} \\   
&+& \frac{1}{3} \overline{\Phi} \left(
\begin{array}{cc}
\Delta & 0 \\       
0 & \Delta^*   
\end{array} \right) S \left(
\begin{array}{cc}
\Delta^* & 0 \\       
0 & \Delta 
\end{array} \right) \Phi + \left(\overline{{\cal N}} \Phi + 
\overline{\Phi} {\cal N} \right)  \label{leff2}
\end{eqnarray}
Here 
\begin{equation} 
S^{-1} = i\fslash{\partial}-m - \Sigma -i\gamma_5 {\bold \tau}\cdot {\bold \pi} \sigma_3 - 
\fslash{\omega} \sigma_3 -i \gamma_5 \left(
\begin{array}{cc}
0 & \Delta_a \beta_a \\     
\Delta_a^* \beta_a & 0 
\end{array} \right), \label{ipr}
\end{equation}
and $S$ in (\ref{leff2}) is the inverse of this.
The meson and diquark fields are now the dynamical variables, while the nucleon field is still an
auxiliary variable.
  
In NM the scalar and vector meson fields have non-zero expectation values, and accordingly 
one shifts the meson 
fields to separate these c-number parts from the fluctuation parts over which one has to integrate.
Since we will describe the nucleon as a quark-diquark state (omitting mesonic fluctuations
for the present), and the NM EOS
in the Hartree approximation, we omit the fluctuation parts of the bosonic fields from now, 
except for the scalar 
diquark field which will be used to construct the nucleon. 
That is, from now $\Sigma$ and $\omega^{\mu}$ will denote only the classical parts of the
scalar and vector fields, which are expressed in terms of the quark fields $\psi$, $\overline{\psi}$ as
\footnote{The quark mass parameter $M$ is considered as a variable in the
effective potential, and its physical value, which is the solution of the gap equation at 
density $\rho>0$ ($\rho=0$), is denoted as $M^*$ ($M_0$).
 The same notation will be used for the
nucleon mass $M_N=M_N(M)$ and the gap $\Delta$, i.e., their physical values are $M_N^*=M_N(M^*)$ and 
$\Delta^*$ at finite density, and $M_{N0}=M_N(M_0)$ and $\Delta_0=0$ at zero density.}
\begin{equation}
\Sigma\equiv -2G_{\pi} 
\langle {\rm NM}|\overline{\psi} \psi|{\rm NM}\rangle \equiv M-m, \,\,\,\,\,\,\,\,\,\,\,\, 
\omega^{\mu} \equiv 2 G_{\omega} \langle{\rm NM}|\overline{\psi}\,\gamma^{\mu} 
\psi|{\rm NM}\rangle \label{class}
\end{equation}
where $|{\rm NM}\rangle$ is the NM ground state. Since the pion field is assumed to have zero expectation
value in NM, it will be omitted from now on.

The integration over the diquark fields can no longer be done exactly. Here we resort to the
stationary phase approximation, where only the terms quadratic in the
 diquark fields are 
retained in the quark
determinant -- i.e., n-point diquark interaction terms
 (n=4,6,8$\dots$) are omitted. 
After performing the Nambu-Gorkov trace in (\ref{leff1}), this amounts to the replacement
\begin{equation}
-\frac{i}{2} {\rm Tr\,\,ln} \,S^{-1} - \int {\rm d}^4x \,\frac{\Delta_a^* \Delta_a}{4 G_s} 
\rightarrow 
-i\, {\rm tr\,\,ln}\, S_0^{-1} + \Delta_a\,D_0^{-1}\, \Delta_a^*. \label{diq}
\end{equation}
Here we introduced the following quantities, which we write down in momentum space for further
use:
\begin{eqnarray}
D_0(k)&=&\frac{-4 G_s}{1+2 G_s \Pi_s(k)} \label{prop0} \\ 
\Pi_s(k) &=& 6i \int \frac{{\rm d}^4 q}{(2\pi)^4} {\rm tr}_D \left[
\gamma_5 S_0(q) \gamma_5 \tilde{S}_0(q-k)\right] \label{pis} \\
S_0(k) &=& \frac{1}{\fslash{k}-M-\fslash{\omega}}, \,\,\,\,\,\,\,\,\,\,\,\,\,\,\,\,\,\,\,\,\,\,
\tilde{S}_0(k) = \frac{1}{\fslash{k}-M+\fslash{\omega}}. \label{s0} 
\end{eqnarray}

Since we wish to construct the nucleon as a quark-diquark state, 
we will retain 
the diquark fields to all orders in the first term of (\ref{leff2}) 
in order to derive the nucleon propagator in the 
ladder approximation to the Faddeev equation. To evaluate this term, we need the form of $S$, which is easily 
derived by inverting (\ref{ipr}) -- see Appendix A. 
Since only terms with the same number of $\Delta$'s and $\Delta^*$'s 
survive after the diquark integration, because of baryon number 
conservation, the non-diagonal parts of 
$S$ in Nambu-Gorkov
space, which involve unpaired fields $\Delta$ or $\Delta^*$, do not contribute. 
After performing the Nambu-Gorkov trace, we therefore find that the first term in (\ref{leff2})
becomes effectively (see Appendix A)
\begin{equation}
\frac{1}{3} \overline{\phi} \left(\Delta \frac{S_0}{1+S_0 \gamma_5 \Delta_{a'} \beta_{a'} \tilde{S}_0 \gamma_5 
\Delta_a^* \beta_a} \Delta^*\right) \phi \label{term}
\end{equation}
After the integration over the diquark fields the nucleon field becomes a dynamical variable,
and this term gives the Faddeev propagator. In order to simplify the formulae, we will
refer here only to the ``static approximation'' to the Faddeev equation \cite{STAT}, which will be
used for the numerical calculations in this paper. 
It is obtained by replacing $\tilde{S}_0$, 
which appears between the diquark fields in the denominator 
of (\ref{term})
and describes the quark exchange between the diquark and the spectator 
quark, according to 
${\tilde S}_0 \rightarrow -1/M$. \footnote{Only in this approximation is the nucleon auxiliary 
field a local one.}    

The integration over the diquark fields is performed as usual by treating the sum of the 
second term in (\ref{diq}) and the term in (\ref{term}) proportional to $\Delta \Delta^*$  
as the free diquark action 
\begin{equation}
{\cal S}_{D0} \equiv \Delta_a D^{-1} \Delta^*_a \equiv
\Delta_a \left(D_0^{-1} + \frac{1}{3} \overline{\phi}  
S_0  \phi \right)\Delta_a^*, \label{dp}
\end{equation}
and the remaining part of (\ref{term}) as the interaction part
\begin{eqnarray}
{\cal S}_{DI}(\Delta,\Delta^*) = \frac{1}{3} \overline{\phi}\left(\Delta  
S_0 \frac{\frac{1}{M} \left(\Delta_{a'} \beta_{a'}\right)\left(\Delta^*_a \beta_a\right)}
{1-\frac{1}{M} S_0 \left(\Delta_{a'} \beta_{a'}\right)\left(\Delta^*_a \beta_a\right)}
S_0  \Delta^* \right) \phi. \label{lint}
\end{eqnarray}
After introducing an auxiliary source term $\Delta^*_a J_a + J^*_a \Delta_a$ into the Lagrangian 
density to perform
the diquark integration, we obtain, with ${\cal S}_D \equiv {\cal S}_{D0}+{\cal S}_{DI}$: 
\begin{eqnarray}
\lefteqn{\int {\cal D} \Delta \int {\cal D} \Delta^* {\rm exp}\left(i{\cal S}_{D}\right)} 
\nonumber \\
& & = {\rm exp}\left(-{\rm Tr}\,{\rm ln}\, D_0^{-1}\right)
{\rm exp}\left(- {\rm Tr}\,{\rm ln}\left[1+\frac{1}{3}D_0 \overline{\phi}
S_0  \phi \right]\right) \label{dint1} \\
& & \times \left[{\rm exp} \left(i{\cal S}_{DI}\left(\frac{\delta}{\delta(iJ^*)},
\frac{\delta}{\delta(iJ)}\right)\right)\,\,{\rm exp}\left(-i J^*_a \,D\, J_a\right)
\right]_{J=J^*=0} \label{dint2}
\end{eqnarray}     
Expanding this expression in powers of the nucleon sources gives the nucleon
propagator as the coefficient of ${\overline \phi}\phi$, and interactions between nucleons
as the higher order terms \footnote{An interesting example is the second exponential factor
in (\ref{dint1}), which describes interactions between nucleons proceeding via
quark exchange processes.}. Here we concentrate
on the nucleon kinetic term. Since the interaction part, Eq.(\ref{lint}), contains
all powers of the diquark fields, many contractions emerge when the derivatives
indicated in (\ref{dint2}) are carried out, even if we restrict ourselves to the nucleon
kinetic term
$\propto {\overline \phi}\phi$. A subset of these contractions constitutes
the ladder approximation to the Faddeev equation (see Appendix A), which leads to 
\begin{eqnarray}
\lefteqn{\int {\cal D} \Delta \int {\cal D} \Delta^* {\rm exp}\left(iS_{D}\right)} \nonumber \\
& &= {\rm exp}\left(-{\rm Tr}\,{\rm ln} \,D_0^{-1}\right)
{\rm exp}\left(i\,\overline{\phi}\, G_{N}\, \phi\right) + {\cal O}
\left[\left(\overline{\phi}\phi\right)^2\right], \label{nprop}
\end{eqnarray}
where the nucleon (quark-diquark) propagator $G_{N}$ in the ladder approximation is expressed in terms of the 
quark-diquark t-matrix ($T_{N}$) in momentum space by
\begin{eqnarray}
G_{N}(p) &=& \Pi_N(p) + \Pi_N(p) T_N(p) \Pi_N(p) \label{gn} \\ 
\Pi_N(p) &=& i \int \frac{{\rm d}^4 k}{(2\pi)^4} S_0(k) D_0(p-k) \label{bubn} \\
T_N(p)&=& \frac{-3}{M} \frac{1}{1+ \frac{3}{M}  \Pi_N(p)} \label{tn} 
\end{eqnarray}
The quark and diquark propagators, $D_0$ and $S_0$, 
appearing in the quark-diquark bubble
graph, (\ref{bubn}), have been given in Eqs.(\ref{prop0}) and (\ref{s0}). 
\footnote{The diquark propagator, $D_0$, is related to the 
quantity $\tau_s$ of Ref.\cite{BT} by $D_0 = i \tau_s$.
The t-matrix (\ref{tn}) is related to $T(p)$ in Eq.(2.7) of 
Ref.\cite{BT} by $T_N(p)=-T(p)$.
We also note that the factor $2G_{\omega}$
in Eq.(\ref{od}) of the present paper was not included in the definition of the field ${\omega}^{\mu}$ in 
Ref.\cite{BT}.} 
As explained in sect.2.1.1 of Ref.\cite{BT}, $T_N$ has a pole at 
$\fslash{p}_N = M_N$, where $p_N^{\mu}=p^{\mu}
-3\omega^{\mu}$. If the total system under consideration (nuclear matter) is at rest,
the space components of the vector field vanish, and the nucleon positive energy spectrum 
becomes $\epsilon_N(p)=E_N(p) + 3\omega^0$, with $E_N(p)=\sqrt{{\bold p}^2+M_N^2}$. 

Inserting these results obtained above into the effective action of Eq.(\ref{leff1}), (\ref{leff2}), and finally
integrating over the nucleon sources (see Appendix A), we obtain the effective action for the 
nucleons in the mean field (Hartree) approximation: 
\begin{eqnarray}
{\cal S}_{\rm eff} &=& -i\left({\rm Tr\,ln}\, S_0^{-1} - {\rm Tr\,ln}\, D_0^{-1} - {\rm Tr\,ln}\, G_N^{-1}\right)
\label{trlog} \\
&+& \overline{N}\, G_{N}^{-1}\, N + \int {\rm d}^4x \left(   
- \frac{\left(M-m\right)^2}{4 G_{\pi}} + \frac{\omega_0^2}{4 G_{\omega}}\right). \label{ln} 
\end{eqnarray}
The three ``trace-log terms'' in (\ref{trlog}) are the results of the integrations over
the quark fields, the diquark fields, and the nucleon sources, respectively. The first
term in (\ref{ln}) is the kinetic term of the nucleon, and the rest are the c-number terms associated
with the mean scalar and vector fields. 

We now introduce a chemical potential ($\mu$) for the nucleon into the kinetic term of (\ref{ln}).
Before that, however, we note that if one had continued and 
integrated ${\rm exp}\left(i{\cal S}_{\rm eff}\right)$ over the nucleon fields as well, 
the last term in (\ref{trlog}) would cancel with the
result of the integration, and one would get an expression which could be obtained directly 
from (\ref{lag}) without any
reference to nucleons. This is obvious, since we introduced the nucleon auxiliary fields
in (\ref{add}) in order that the generating functional would be unchanged. This observation is 
important, however, because
it means that, as a consequence of the composite nature of the nucleon, there are no 
nucleon vacuum loop terms.
That is, there are no nucleon loop contributions to the 
effective potential {\it unless} we 
introduce a chemical potential
into the kinetic energy term of (\ref{ln}). 
The vacuum loop terms are exhausted already by the 
quark and diquark loops in (\ref{trlog}), 
which describe the
polarization of the quark Dirac sea and the effects of quark-quark correlations (so called, 
ring contributions) in the quark Dirac sea, respectively. 

Concerning the ring contributions (corresponding to the second term in Eq.(\ref{trlog})), we note that meson 
loops, which were omitted from the outset in our simplified treatment, would give very 
similar terms. A consistent treatment of the ring contributions would therefore require
both meson and diquark loops. This is beyond the scope of the present
approximation, where we have retained the diquark fields solely in order to construct 
the single nucleon as a quark-diquark state. In the framework of the Hartree approximation
for the EOS, we therefore drop the second term in Eq.(\ref{trlog}).

One can rewrite the nucleon kinetic term by introducing a renormalized nucleon field ($\hat{N}$), such that, near the pole, 
the renormalized propagator
($\hat{G}_N$) behaves as $\hat{G}_N \rightarrow \fslash{p}_N-M_N$. 
\footnote{Mathematically, we write $N=\sqrt{Z_N} \hat{N}$ and 
$G_N=Z_N \hat{G_N}$ with 
$Z_N=-(M/3) \left[\left(\partial \Pi_N(p_N)/\partial \fslash{p}_N\right)_
{\fslash{p}_N=M_N}\right]^{-1}$.} 
The pole part of the nucleon kinetic term including the chemical potential then becomes
\begin{eqnarray}
\overline{N}\,\ G_{N}^{-1}\, N \rightarrow
\overline{\hat{N}} \left(\fslash{p}-M_N + \mu^*\gamma^0\right) \hat{N},  \label{kn}
\end{eqnarray}
where $\mu^* \equiv \mu-3\,\omega^0$. 

To continue the derivation of the effective potential ($V$) for NM in the path integral 
formalism, we finally have to integrate over the nucleon field and express the result as 
$S_{eff}=- \int {\rm d}^4x V$, where $V$ is a function of the classical 
fields $M$ and $\omega^0$.  
Combining the result of the integration of 
the kinetic term (\ref{kn}) with the last term of Eq.(\ref{trlog}), we get the nucleonic contribution  
to the effective potential\footnote{For the explicit evaluation of the Dirac determinant, 
see Appendix B for the case $\Delta=0$.
The trace indicated by the symbol tr does not include the functional trace.
The subtraction of the $\mu=0$ term in (\ref{vn}) arises from the last term in (\ref{trlog}), and
the factor 2 in the second line comes from the isospin trace.}
\begin{eqnarray}
\lefteqn{V_N = i\, \int \frac{{\rm d}^4 p}{(2\pi)^4} {\rm tr\,ln} \, 
\left(\fslash{p}-M_N + \mu^*\gamma^0\right) - \left(\mu=0\,\,{\rm term}\right)}
\label{vn} \\
&=& \frac{i}{2} \times 2 \int \frac{{\rm d}^4 p}{(2\pi)^4} \left[{\rm tr_D\,ln} 
\left(-\fslash{p}-M_N + \mu^*\gamma^0\right)
\left(\fslash{p}-M_N + \mu^*\gamma^0\right)\right] - \left(\mu=0\,\,{\rm term}\right) \nonumber \\
&=& 2i \int \frac{{\rm d}^4 p}{(2\pi)^4} \left[{\rm ln}\, 
\frac{p_0^2-(E_N(p)+\mu^*)^2}{p_0^2-E_N(p)^2}
+ {\rm ln} \,\frac{p_0^2-(E_N(p)-\mu^*)^2}{p_0^2-E_N(p)^2} \right] - 
\left(\mu^* \rightarrow -3\omega^0\right) \nonumber \\
\label{vn1}  \\
&=& -4 \int \frac{{\rm d}^3 p}{(2\pi)^3} \Theta\left(\mu^* - E_N(p)\right) \cdot \left(\mu^*-E_N(p)\right), \label{fermi}
\end{eqnarray}
which is nothing but the familiar contribution to the effective potential arising from the 
nucleon Fermi motion. Adding the quark loop
contribution (the first 
term in (\ref{trlog})) and the mean field terms of (\ref{ln}), 
we finally obtain the effective potential for NM in the form
\begin{equation}
V = V_{\rm vac} + V_N - \frac{\omega_0^2}{4G_{\omega}} \label{vtot}    
\end{equation}
where $V_N$ is given by (\ref{fermi}), and the vacuum term is
\begin{equation}
V_{\rm vac} = 12\, i \int \frac{{\rm d}^4 k}{(2\pi)^4} \,{\rm ln}\, \frac{k^2-M^2}{k^2-M_0^2} + \frac{\left(M-m\right)^2}{4G_{\pi}}
- \frac{\left(M_0-m\right)^2}{4G_{\pi}}. \label{vac}
\end{equation}
Here we subtracted the zero density contribution ($M=M_0$). The values of $\omega_0$ and $M$ for fixed $\mu$ are
determined by the equations\footnote{In principle, the quantity which has to be minimized is the effective potential
after Wick rotation. In this case the last term in Eq.(\ref{vtot}) changes its sign, and the physical value of
$\omega_0$ is a minimum of $V$.} $\partial V/\partial \omega_=\partial V/\partial M=0$.  
The pressure as a function of $\mu$ is then obtained as $P=-V$. The baryon density follows from $\rho=- \partial V/\partial \mu$,    
and the energy density is given by ${\cal E}(\rho)=V+\mu \rho$. 

The effective potential (\ref{vtot}) differs from the familiar expression in the linear sigma model
for point-nucleons \cite{SIG} only through the dependence of the nucleon mass 
$M_N$ on the mean scalar field (or, equivalently, on $M$).  
Since the scalar field now couples to the quarks in the nucleon instead of an elementary nucleon, 
the function $M_N(M)$, 
which is determined by the pole position of the quark-diquark t-matrix (\ref{tn}), is in
general a non-linear function of $M$. If it has a large positive curvature 
(scalar polarizability), the binding energy  
per nucleon saturates, as has been discussed in detail in Ref.\cite{BT}.  

\subsection{Color superconducting quark matter}

To obtain the EOS of QM in the Hartree approximation, we introduce a chemical potential for quark number
$\mu_q=\mu/3$ into the Lagrangian density (\ref{lag}), and assume that the fields $\Sigma$, 
$\omega^0$ and $\Delta_{11}$
have finite expectation values. The choice $D=1$ for the diquark field $\Delta_{Da}$ corresponds to a special
choice of the phase and breaks the $U(1)$ symmetry, while the choice $a=1$ breaks the color $SU(3)$ symmetry down to
$SU(2)$ \footnote{There are 5 different Goldstone bosons corresponding to 
these broken symmetries. 
If we start from the 
($\sigma_1\,\beta_1$) mode ($\Delta_{11}$), which characterizes the ground state, the $U(1)$ phase 
rotation mixes in the ($\sigma_2\,\beta_1$) mode,
and the color $SU(3)$ rotation mixes in the ($\sigma_1\,\beta_2$), ($\sigma_1\,\beta_3$) and ($\sigma_2\,\beta_a$)
($a=1,2,3$) modes, which correspond to the Goldstone bosons. Note that the ($\sigma_2\,\beta_1$) mode is mixed into
the ground state by both 
phase and color rotations. The quark-quark t-matrices in these channels have zero energy poles. For comparison, we
note that the familiar Goldstone pion in the present Nambu-Gorkov
formalism emerges as a $\sigma_3$-mode in the ($0^-, T=1$) color singlet channel, which is mixed into the 
$\sigma_1$-mode in the ($0^+, T=0$) color singlet channel of the ground state by the chiral rotation.}.
Since we do not consider fluctuations of the bosonic fields in this work, it is sufficient to use
the following Lagrangian density:
\begin{equation}
{\cal L} = \overline{\Psi} S^{-1} \Psi - \frac{\left(M-m\right)^2}{4 G_{\pi}} + \frac{\omega_0^2}{4 G_{\omega}}
- \frac{\hat{\Delta}^2}{4 G_s}.  
\label{lmfq}
\end{equation}
Here
\begin{equation} 
S^{-1} = i\fslash{\partial}-M +\mu^*_q \,\gamma^0\, \sigma_3 -i \gamma_5\, \sigma_1 \,
{\hat \Delta}\, \beta_1,
\label{iprq}
\end{equation}
where $\mu_q^*=\mu_q-\omega_0$, and the meson mean fields are defined in terms of the
ordinary quark fields ${\overline \psi}$, $\psi$ by (compare with Eqs. (\ref{sp}), (\ref{od}), 
and (\ref{class}))
\begin{eqnarray}
M &=& m-2G_{\pi} <{\rm QM}|{\overline \psi}\psi|{\rm QM}>,
\,\,\,\,\,\,\,\,\,\,\,\,\,\,\,\,\,\,\,\,
\omega_0 = 2 G_{\omega} \langle {\rm QM}| \psi^{\dagger} \psi|{\rm QM} \rangle, 
\nonumber \\
{\hat \Delta}&=&-2G_s \frac{1}{2} \langle{\rm QM}|\left({\overline \psi}i\gamma_5 \beta_1
C\tau_2 {\overline \psi}^T - \psi^T C^{-1}\tau_2 i \gamma_5 \beta_1 \psi \right)
|{\rm QM}\rangle . \label{mfqm}
\end{eqnarray}

The poles of the quark propagator, $S$, are easily be determined by 
inverting Eq.(\ref{iprq}). 
There are 4 poles, at $p_0= \pm \sqrt{(E_q(p)\pm \mu^*)^2+\Delta^2}$, each with degeneracy 8, 
and 4 poles at $p_0=\pm|E_q(p)\pm \mu^*|$, each with degeneracy 4. Here $E_q(p)=
\sqrt{{\bold p}^2+M^2}$, and 
${\displaystyle \Delta \equiv \sqrt{\frac{3}{2}} \hat{\Delta}}$. 

Integration over the quark fields gives a contribution ${\displaystyle -\frac{i}{2} {\rm Tr\,ln}\, S^{-1}}$ to the effective action,
which can be evaluated (Appendix B) with the result
\begin{eqnarray}
\lefteqn{- \frac{i}{2} {\rm Tr\,ln}\, S^{-1} = 
- \int {\rm d}^4 x} \nonumber \\
& & \times
2i \int \frac{{\rm d}^4 p}{(2\pi)^4} \left[2\, {\rm ln} \left(p_0^2-(E_q(p)+\mu_q^*)^2 -
\Delta^2 \right) + 2 \,{\rm ln} \left(p_0^2-(E_q(p)-\mu_q^*)^2 - \Delta^2 \right) \right. 
\nonumber \\
& & \left. + {\rm ln} \left(p_0^2-(E_p+\mu_q^*)^2 \right) +  
{\rm ln} \left(p_0^2-(E_p-\mu_q^*)^2 \right) \right].
\label{qdet}
\end{eqnarray}
If we separate the contributions surviving for $\Delta=0$, we finally obtain the following form of the
effective potential for QM (compare to (\ref{vtot}) for the NM case):
\begin{equation}
V = V_{\rm vac} + V_Q + V_{\Delta} -\frac{\omega_0^2}{4 G_{\omega}} \label{vq}
\end{equation}
where $V_{\rm vac}$ has the same form as in nuclear matter (see Eq.(\ref{vac})), $V_Q$ 
is the contribution of the quark Fermi motion given by
\begin{equation}    
V_Q = -12 \int \frac{{\rm d}^3 p}{(2\pi)^3} \Theta\left(\mu_q^* - E_q(p)\right)
\left(\mu_q^*-E_q(p)\right), \label{fermiq}
\end{equation}
and $V_{\Delta}$ is the contribution arising from the finite gap: 
\begin{eqnarray}
V_{\Delta} &=& 4i
\int \frac{{\rm d}^4 p}{(2\pi)^4} \left[ {\rm ln} \frac{p_0^2-(E_q(p)+\mu_q^*)^2 -\Delta^2}
{p_0^2-(E_q(p)+\mu_q^*)^2} + {\rm ln} \frac{p_0^2-(E_q(p)-\mu_q^*)^2 -\Delta^2}
{p_0^2-(E_q(p)-\mu_q^*)^2}\right] \nonumber \\ 
&+& \frac{\Delta^2}{6 G_s}. \label{vd}
\end{eqnarray}
The values of $\omega_0$, $M$ and $\Delta$ for fixed chemical potential are obtained by solving the
equations $\partial V/\partial \omega_0=\partial V/\partial M=\partial V/\partial \Delta=0$. (Concerning
the dependence on $\omega_0$, see the last footnote in subsect.3.1.) 
The pressure, baryon density and energy density are then obtained from 
${\displaystyle P=-V, \,\, 
\rho=-\frac{\partial V}{\partial \mu}}$, and ${\cal E} = V+\mu \rho$, where $\mu=3 \mu_q$
is the chemical potential for baryon number as before.

\section{Numerical results}
In the numerical calculations discussed below, the proper time regularization scheme will be used. In this scheme,
one performs a Wick rotation of the energy variable of the loop integral, and then introduces the
replacements
\begin{eqnarray}
{\rm ln\, A} \rightarrow \int_{1/\Lambda_{\rm UV}^2}^{1/\Lambda_{\rm IR}^2} \frac{{\rm d}\tau}{\tau}
e^{-\tau A} , \,\,\,\,\,\,\,\,\,\,
\frac{1}{A^n} \rightarrow \frac{1}{(n-1)!} \int_{1/\Lambda_{\rm UV}^2}^{1/\Lambda_{\rm IR}^2} {\rm d}\tau\, 
\tau^{n-1} e^{-\tau A} \,\,\,\,\,(n\geq 1), \nonumber \\ \label{pt}
\end{eqnarray}
where $A$ depends on the momenta and, possibly, Feynman parameters, and $\Lambda_{\rm UV}$ and
$\Lambda_{\rm IR}$ denote the UV and IR cut-offs, respectively.  

As discussed in detail in Ref.\cite{BT}, the validity of 
the ``static approximation'' to the Faddeev equation, which
leads to the contact-type quark-diquark interaction $\propto 1/M$ in 
the t-matrix (\ref{tn}), breaks down
as $M$ decreases with increasing density, and leads to large 
deviations from the exact Faddeev results for the
function $M_N(M)$.
On the other hand, it was shown that if one fixes the quark-diquark 
interaction to its value at zero
density ($1/M \rightarrow 1/M_0$ in the denominator of Eq.(\ref{tn})), 
the exact Faddeev results are qualitatively
reproduced. Furthermore, if one uses an interpolating form of the 
quark-diquark interaction
($1/M \rightarrow (1/M_0)(M_0+c)/(M+c)$, with $c=0.7$ GeV, 
in the denominator of (\ref{tn})), 
the agreement is
very good. We will use this interpolating form in the following calculations.   

\subsection{EOS of NM}
In order to provide a basis for our later discussions on the NM$\rightarrow$ QM phase transition,
we first reproduce in Figs. 1 and 2 the results obtained in Ref.\cite{BT} for the nucleon mass as a function 
of the scalar potential, $M-M_0$, as well as the binding energy per nucleon as a function of density.  
The parameters used in this calculation are shown in Table 1.
For the choice $\Lambda_{IR}=0$ there is an unphysical threshold for the decay of the nucleon into a quark and a diquark,
and the function $M_N(M)$ cannot develop a sufficiently large curvature (scalar polarizability), which 
is required to stabilize the system. In contrast, for the case $\Lambda_{IR}=0.2$ GeV the nucleon 
mass exceeds
the would-be threshold for large scalar potentials, leading to a sufficiently large scalar polarizability and
to saturation of the binding energy \footnote{We recall from Ref.\cite{BT} that the only free
parameter for the calculation of the NM EOS, $r_{\omega}$, has been adjusted so that the binding
energy curve passes through the empirical saturation point. That is, with the limited number of
parameters in this simple NJL model, it is
not possible to ensure that the calculated saturation point agrees with the empirical one.}.     
The effective masses of the nucleon and the quark 
for the case $\Lambda_{\rm IR}=0.2$ GeV are shown by the solid lines in Fig. 3. We observe
that there is only a limited tendency toward chiral restoration in NM.  
The pressure of NM will be discussed later in connection with the phase transition to QM. 

Let us discuss here another aspect which provides a further illustration of the importance of
avoiding unphysical decay 
thresholds, namely the mass of the $\omega$-meson in NM. We take the simplest
case of an $\omega$-meson at rest (${\bold q}=0$), and consider the pole position $q_0\equiv M_{\omega}^*$ of the 
effective NN interaction in the t-channel with quantum numbers $1^+$. The effective NN interaction for the case 
${\bold q}=q_0=0$ is the familiar Landau-Migdal interaction, which was derived in 
Ref.\cite{BT} directly from the
expression for the NM energy density. The part associated with the spatial 
components of the exchanged $\omega$-meson has the form
\footnote{See Eq.(2.50) of Ref.\cite{BT}. We note that the
factor $\frac{1}{2}$ appearing there in the denominator is a misprint. 
The correct expression
is (\ref{lm}).}
\begin{eqnarray}
f_{\omega}({\bold p}',{\bold p})=- \frac{18\,G_{\omega}}{1+18\,G_{\omega} \frac{\rho}{E_F^*}}
\frac{{\bold p}'\cdot{\bold p}}{E^*_N(p')E^*_N(p')},  \label{lm}
\end{eqnarray}
where $E_F^*=\sqrt{p_F^2+M_N^{*2}}$. 
The part involving $18\,G_{\omega} \frac{\rho}{E_F^*}$ 
is a density dependent N-loop contribution -- i.e., the Fermi average over 
the Z-graph for external vector fields with zero momentum. 
(This is the same type of 
medium correction which reduces the enhanced isoscalar 
magnetic moments in relativistic
many-nucleon theories \cite{MM}.) 

Since the nucleon remains a rather massive object
even at finite density (see Fig. 3), we can expect that for the case of finite energy 
transfer this N-loop contribution will not depend very strongly 
on $q_0$, and we therefore approximate it here by its value at $q_0=0$. Then, for finite $q_0$, 
the only additional
contribution arises from the exchange of $q{\overline q}$ pairs in the t-channel, described by the 
transverse part of the polarization (bubble graph)
$\Pi_{\omega}$ -- see Appendix C. The constant in the denominator 
of Eq.(\ref{lm}) is then 
replaced according to
\begin{equation}
18\,G_{\omega} \frac{\rho}{E_F^*} \longrightarrow 18\,G_{\omega} \frac{\rho}{E_F^*}+2G_{\omega} \Pi_{\omega}(q_0).
\label{denom} 
\end{equation}
The two terms on the r.h.s. of Eq.(\ref{denom}) work in opposite directions for increasing density; 
the N-loop term represents 
the effect of the Pauli exclusion principle acting on the nucleons and tends to make the $\omega$-meson heavier, while the
$q{\overline q}$ excitation piece tends to make the $\omega$-meson lighter (since the mass of 
the $q{\overline q}$ pair is reduced).

Using the density dependent quark mass in NM, as shown in Fig. 3, and the 
value of $r_{\omega}$ 
from Table 1, we obtain the result for $M_{\omega}^*$  
shown in Fig. 4. First we note that $r_{\omega}$ has been adjusted in Ref.\cite{BT} 
so that the binding energy curve
passes through the empirical saturation point. From Fig. 4 we see that this leads to an 
$\omega$ meson mass of 827 MeV at zero
density, which is larger than the experimental value of 783 MeV, but nevertheless reasonable. 
Second, the $\omega$ meson mass 
decreases by about 80 MeV at normal NM density and by about 120 MeV for 
$\rho\simeq$ 0.5 fm$^{-3}$. For higher
densities it increases slightly because of the increasing importance of Pauli blocking
\footnote{In view of the interest in searching for possible bound $\omega$-nucleus states \cite{Hayano:1998sy}, it is
interesting to note that these results are very close to those found in the quark meson
coupling model \cite{Tsushima:1998qw}, with the $\omega$-meson being somewhat less bound than it would be
in Walecka-type models \cite{Saito:1998ev}.}.
In comparison with our earlier NJL model calculations, we note that the overall picture 
for the $\omega$-meson mass is similar to that for the $\sigma$-meson mass 
(see Fig. 11 of Ref.\cite{BT}), 
and also to the 
nucleon and quark masses shown in Fig. 3. That is, the density dependence is rather 
mild, and there are no sudden changes of the behavior at high density. 
Third, we note
that the $\omega$ meson mass shown in Fig. 4 is always above the 
would-be $q{\overline q}$ threshold.
If we set $\Lambda_{\rm IR}=0$, leaving the other parameters unchanged, we 
would still get a bound
state at zero density ($M_{\omega}=780$ MeV), but this state would already
dissolve at $\rho\simeq 0.04$ fm$^{^3}$, because of the decreasing $q{\overline q}$
threshold. Once again this shows the importance of avoiding the unphysical threshold in the NM
calculation.  

\subsection{EOS of QM without the effect of diquark condensation}
The effective quark mass in QM is shown as a function of the density by the
dashed line in Fig. 3. The parameters used in this calculation are the same as for NM, see
Table 1. (We note that the value of $r_{\omega}$ has no influence on $M^*$.) 
The sharp decrease of $M^*$ in QM
around 0.3 fm$^{-3}$ reflects the well known chiral phase transition 
which would happen 
for the case $m=0$\footnote{Since in the numerical calculations of 
this paper we always adjust
the parameters so as to reproduce the experimental pion mass, we refer to this kind of
behavior as the ``would-be chiral phase transition'', or simply 
the ``chiral phase transition''.}. 
We see that the NM and QM cases
are qualitatively very different and one should not treat 
ordinary finite density matter as QM.

Fig. 5 shows the variation of the pressure as a function of the 
chemical potential for NM 
(solid line), and for QM
with the same model parameters as for NM (dashed line labeled by $r_{\omega}=0.37$), 
as well as for the choice $r_{\omega}=0$
(dashed line labeled by $r_{\omega}=0$). The case of zero density 
corresponds to the point $(P,\mu)=(0, M_{N0})$ for NM, and to $(0, 3M_{0})$ for QM, where
$M_{N0}=940$ MeV and $M_0=400$ MeV are the zero density values of the nucleon and 
constituent quark masses.
Starting from this point, the density increases along the lines. The vacuum solution $P=0$, which corresponds to
a minimum of the effective potential for $\mu<M_{N0}$ in NM, and for $\mu<3M_{0}$ in QM, 
is not shown in this plot. 
For a qualitative understanding of our following discussions it is helpful to recall that attractive effects,
which soften the EOS, make the $P(\mu)$ curves steeper, and repulsive effects make them flatter.   

Let us first summarize some well known points concerning the phase transitions which occur 
within the NM and QM phases separately. 
In the region below the saturation density, the NM EOS shows the familiar behavior of a first order 
gas-liquid phase transition, which is shown in more detail in the insert of Fig. 5.
The pressure decreases with increasing density until it reaches a minimum at some density $\rho_c$, and then it increases and 
passes through $P=0$ at the saturation density $\rho_0$. In other words, for densities between 
$\rho_c$ and $\rho_0$ there are 3 extrema of the effective potential as a function of $M$, corresponding to 
(i) a maximum at negative
pressure corresponding to unstable NM droplets, (ii) the
NM phase, which has negative pressure for $\rho<\rho_0$ (quasi-stable NM droplets), and positive pressure 
for $\rho>\rho_0$ (stable NM), and (iii) the vacuum
phase ($P=0$), which is the ground state (state with the largest pressure) for $\rho<\rho_0$. 

The EOS of QM for the case
$r_{\omega}=0$ has a bound state, and in this case the behavior is qualitatively similar to that of 
NM, discussed above. The gas-liquid phase transition of NM now corresponds to the familiar first
order chiral phase transition in QM, where the unstable branch with decreasing pressure 
is characterized by massive
constituent quarks (broken chiral symmetry), and the metastable ($P<0$) and stable ($P>0$)
branches with increasing pressure correspond to the chirally restored
phase. The first order phase transition occurs
at $P=0$ in this case, but if one increases $r_{\omega}$ the curve will cross itself at 
finite $P$, and a further increase of $r_{\omega}$ leads to a chiral phase transition of second 
order, as shown for $r_{\omega}=0.37$
in Fig. 5. In this case there is only one minimum of the effective potential as a function of $M$ 
for any given $\mu$, and at some $\mu$ the maximum at $M \simeq 0$ turns into a minimum.
The density increases along the dashed line for $r_{\omega}=0.37$, starting at the point 
where $\mu=3\,M_0=1200$ MeV and $P=0$, and at $\mu=1500$ MeV ($\rho=0.41$ fm$^{-3}$) the quark 
effective mass is already as small as $M^*=53$ MeV.    

The branch of the QM EOS corresponding to (almost) massless quarks is essentially the 
EOS of a free massless quark gas \cite{PT}.
This is demonstrated by the dotted lines in Fig.5, which show the EOS for a massless quark gas 
for two choices of the
bag constant $B$: $B_{\rm NJL}=139.7$ MeV\,fm$^{-3}$ is the height of the Mexican hat
vacuum potential (\ref{vac}) at $M=0$, and $B_{\rm MIT}=57.5$ MeV\,fm$^{-3}$ is the value 
used in the MIT bag model. By comparing
the dashed line for $r_{\omega}=0$ with the dotted line for $B=B_{\rm NJL}$, we see that,
in the region which is relevant for a discussion of the NM$\rightarrow$ QM phase transition, 
one could neglect
the spontaneous breaking of chiral symmetry from the outset and start with massless quarks. 

By comparing the solid line with the dashed line for the case $r_{\omega}=0.37$ in Fig. 5, we see
that there is no NM $\rightarrow$ QM phase transition if we use the
same model parameters to describe the two phases. That is, NM would be the ground state for 
all densities. It will be demonstrated later that this unsatisfactory situation cannot be 
improved by including the possibility of diquark condensation. (A stiffer EOS for NM would certainly come
closer to the QM result at high densities, but one never gets a crossing for any reasonable 
NM EOS.) On the other hand, the dashed curve with $r_{\omega}=0$ is much closer to the
NM curve, although still there is no phase transition. However, it will be shown below that
this situation changes immediately if one allows for the possibility of diquark
condensation. \footnote{We also note that the dotted line labeled by $B=B_{\rm MIT}$ in Fig. 5,
which has only little to do with the NJL model, 
crosses the NM line in the high density region, but there is also an unphysical crossing
in the low density region.} We conclude from Fig. 5 that it is not
possible to get a phase transition from NM to normal QM by using the same parameters for
the description of the two phases,
in particular the same strength of the vector interaction. This point was also noted in a recent
investigation \cite{REC}. However, if one uses
$r_{\omega}=0$ in QM and further considers effects which soften the QM EOS, like diquark 
or pion condensation, there could be a phase transition. 

Based on the preceding discussion, 
we are led to consider the question of whether it is reasonable to use the same model parameters 
for NM and QM or not. Two among the
parameters listed in the second column of Table 1 have been fixed by considering only the properties of NM, namely
$\Lambda_{\rm IR}$ and $r_{\omega}$. First, one needs a finite IR cut-off in order 
to describe a stable
NM state, but the results are rather insensitive to its actual value as long as 
$\Lambda_{\rm IR}>0.1$ GeV.
The introduction of a finite $\Lambda_{\rm IR}$ takes into account one
aspect of confinement physics, which is important for single hadrons and NM.  However, 
there is no reason a priori to introduce an IR cut-off in QM. On the
contrary, one might expect that quark decay processes are welcome in this case, and 
that the choice $\Lambda_{\rm IR}=0$ would be more reasonable.  
Second, one also needs a finite value of $r_{\omega}$ in NM in order to 
reproduce the
saturation properties. Once again, there is no compelling reason to use 
the same value in QM, and from Fig. 5 we see
that the QM EOS depends very sensitively on the actual value of $r_{\omega}$.

In order to shed some light on the question of which values of $\Lambda_{\rm IR}$ and 
$r_{\omega}$ are reasonable 
in QM, we return to the problem of the $\omega$ meson mass as an example, and investigate whether 
or not there is a pole corresponding to the $\omega$ meson in the 
$q{\overline q}$ t-matrix in QM. Contrary to the NM case discussed in the previous subsection, 
we now have a chemical potential for the
quarks, which leads to a modification of the $q{\overline q}$ bubble graph arising from Pauli blocking. 
(The detailed formulae are given in Appendix C.) Fig. 6 shows the 
situation for the choice 
$\Lambda_{\rm IR}=0.2$ GeV in QM. The
would-be threshold, which is $2\sqrt{p_F^2+M^{*2}}$, is shown as a function of density by the dashed line,
and the pole positions for ${\bold q}=0$ are shown for various values of $r_{\omega}$ by the solid lines
\footnote{The density dependence shown by the dashed line reflects the behavior of the chemical potential
$\mu_q^*$, see the dashed line for $r_{\omega}=0$ in Fig. 5. The decrease of $M^*$ with 
increasing density 
lowers the threshold, while the effect of Pauli blocking raises the threshold. In the low density region the first
effect dominates, while in the high density region, where the quarks are practically massless, the Pauli 
effect leads to a continuous increase of the threshold. Figs. 6 and 7 show that the density dependence of the
pole positions follows the behavior of the dashes lines}. 
We see that poles exist even for very small
values of the coupling constant. We expect that the existence of a hadron pole in QM for all 
densities is not restricted to the case of the $\omega$ meson shown here, but can be ascribed to the finite
IR cut-off. This observation indicates that it is not reasonable to use a finite IR cut-off in QM. 
(Another argument in favor of the choice $\Lambda_{\rm IR}=0$ will be given in the next
subsection.)

Fig. 7 shows the situation for the case $\Lambda_{\rm IR}=0$ in QM. We see that for the 
choice $r_{\omega}=0.37$ there is again a pole for all densities, which now is below the threshold. If the value of 
$r_{\omega}$ is decreased, the pole
disappears in the low density region, but at high densities it re-appears 
because of the increasing 
threshold. For $r_{\omega}=0.17$ the pole exists only for $\rho>0.8$ fm$^{-3}$. Since the 
appearance of a meson
pole in high density QM, which was not present in the vacuum, is physically unreasonable, Fig. 7 indicates 
that one should really set $r_{\omega}=0$ in QM.

Based on these discussions, we will use $\Lambda_{\rm IR}=0$ and $r_{\omega}=0$ in QM 
in our following discussions. The corresponding parameter set is shown in the third column of Table 1.
As compared to the case $\Lambda_{\rm IR}=0.2$ GeV, there are small changes in $m$, $G_{\pi}$ and
$\Lambda_{\rm UV}$ in order to reproduce the same input values $m_{\pi}=140$ MeV, $f_{\pi}=93$ MeV 
and $M_0=400$ MeV. We will investigate the dependence of the QM results on the pairing 
strength, and therefore we treat $r_s$ as a free parameter in QM. (Some numerical results obtained
by using the same values of $\Lambda_{\rm IR}$ and/or $r_{\omega}$ as for NM are shown in Appendix D.)

\subsection{EOS of QM including the effect of diquark condensation}
The effect of diquark condensation on the QM EOS is represented by the piece (\ref{vd}) of the effective
potential. In order to discuss the nature of the phase transition to the color symmetry broken phase,
it is useful to look at the gap equation ${\displaystyle \frac{\partial V}{\partial \Delta}=0}$ for fixed
$M$. Besides the trivial solution $\Delta=0$, the nontrivial solution is obtained by solving the equation
\begin{equation}
1=f(\Delta^2), \label{gap1}
\end{equation}
where the explicit form of the function $f(\Delta^2)$ in the proper 
time regularization scheme
is
\begin{equation}
f(\Delta^2)=\frac{12 G_s}{\pi^3} \int_0^{\infty}{\rm d}k_0 \int_0^{\infty}k^2\,{\rm d}k
\left(F_+ + F_-\right), \label{fd}
\end{equation}
where
\begin{equation}  
F_{\pm} = \frac{{\rm exp}\left(- \frac{k_0^2+\epsilon_{\pm}^2}{\Lambda_{\rm UV}^2}\right) -
{\rm exp}\left(- \frac{k_0^2+\epsilon_{\pm}^2}{\Lambda_{IR}^2}\right)}{k_0^2+\epsilon_{\pm}^2}
\label{fd1}
\end{equation}
with $\epsilon_{\pm}^2=\left(E_q(k) \pm \mu_q^*\right)^2+\Delta^2$. 
It is easy to show that    
$f(\Delta^2)$ is a monotonically decreasing function of $\Delta^2$. For 
the case $\Lambda_{IR}>0$, its value at
$\Delta^2=0$ is finite, and therefore we have one unique nontrivial solution if $f(0)>1$. In other
words, for the case $\Lambda_{IR}>0$ there exists a threshold value of the chemical potential (for fixed 
coupling constant $G_s$), 
or of the coupling constant $G_s$ (for fixed chemical potential), below which we have 
only the trivial solution $\Delta=0$, and above which there is one unique minimum
at $\Delta>0$. The transition to the color broken phase is therefore of
second order in this case. On the
other hand, for $\Lambda_{IR}=0$, the function $f(\Delta^2)$ has a logarithmic 
singularity for
$\Delta^2\rightarrow 0$, and therefore the nontrivial solution exists for all values of the 
chemical potential
and of the coupling constant. In equations, if $\Lambda_{IR}=0$ the behavior for small 
$\Delta \rightarrow 0$ is expressed as follows:
\begin{eqnarray}
f(\Delta^2) &=&- \frac{6 G_s}{\pi^2} p_F \,\mu_q^{*2}\, {\rm ln} \Delta^2 +  \dots \label{b1} \\  
\Delta &\propto& {\rm exp}\left(-\frac{\pi^2}{12 G_s p_F \mu_q^*}\right), \label{b2}
\end{eqnarray}
where $p_F\equiv\sqrt{\mu_q^{*2}-M^2}$.
In this case, $V_{\Delta}$ has the Mexican hat shape for all values of $\mu_q^*$ 
and $G_s$, and the system is
always in the color symmetry broken phase. This behavior, which is very familiar from ordinary BCS theory
\cite{BCS}, can be
derived directly from QCD without using an effective quark theory \cite{DEL}. 
This observation further supports our choice $\Lambda_{\rm IR}=0$ in QM.

In Fig. 8 we show the $P-\mu$ plots. The dotted line ($r_s=0$) is very similar to the corresponding
line ($r_{\omega}=0$) of Fig. 5, and shows that there is no phase transition from NM to normal QM. 
However, the 
softening of the QM EOS arising from diquark condensation leads to a first order phase transition, 
even for relatively small values of $r_s$. For example, for $r_s=0.1$ the phase transition 
appears only at extremely high densities, but for $r_s=0.15$ the phase transition sets in at 
$0.72$ fm$^{-3}$ in NM, and ends at $1.21$ fm$^{-3}$ in QM. The corresponding transition densities for $r_s=0.2$  
are $0.57$ fm$^{-3}$ and $0.95$ fm$^{-3}$. For the case $r_s=0.25$ the transition densities are already somewhat too 
low ($0.42$ fm$^{-3}$ and $0.77$ fm$^{-3}$), and $r_s=0.3$ is definitely too strong since QM would be the ground state
for practically all densities. One should also note that, for any reasonable scenario, those parts of the
QM curves which reflect the chiral phase transition are always below the NM curve. 
The chiral phase
transition in QM is therefore not very relevant for the discussion of the NM $\rightarrow$ QM phase transition,
and one could set $M\equiv 0$ in QM from the beginning with practically identical conclusions.

In Fig. 9 we show the quark effective masses as functions of 
the density for the same 
values of $r_s$ as in Fig. 8, and in Fig. 10 we show the gaps as functions 
of the density. The behaviors of the
curves $M^*(\rho)$ still reflect the chiral phase transition in QM, 
although the density region where
$M^*$ varies rapidly is shifted upward compared to the 
case $r_s=0$. Fig. 10 shows that the 
gaps in the present calculation are
rather large. In the density region where QM becomes the ground 
state the gap is of the order 
of 150 MeV to 250 MeV for those values of $r_s$ which give reasonable transition densities. 
This figure also supports the conclusion
drawn from Fig. 8 that $r_s=0.3$ is already too strong, since the gap 
would exceed 300 MeV 
in the high density region, which is definitely too large. 

With regard to the dependence of the above results on the pairing 
strength, $r_s$, we would like to 
make two remarks. The first one concerns the regularization scheme. If we compare, for example, 
the present proper time scheme to the sharp 3-momentum cut-off scheme, a glance at Table 1 of 
Ref.\cite{BT} shows
that the coupling constant $G_{\pi}$ in the proper time scheme is almost three times as 
large as the one in the
3-momentum cut-off scheme. Therefore, for a given value of $r_s$, 
the actual value of the coupling 
constant, $G_s$, in the 
proper time scheme is much larger than in the 3-momentum cut-off scheme, which partially
explains why our gaps are larger than in previous works \cite{CS3}. One also has to 
note that the most
important contributions to the gap equation (\ref{gap1}) come from the region 
$|{\bold k}| \simeq \mu_q^*$ -- see Eqs.(\ref{fd}) and (\ref{fd1}). 
In the high density region, this is of the same order as 
the UV cut-off in the 3-momentum
scheme, and therefore the peak contributions to the gap equation are cut out artificially. 
On the other hand, the
proper time scheme introduces a smooth cut-off function, and the 
gap can become quite large 
in the high density region.

The second remark is that the value $r_s=0.51$, obtained by fitting the nucleon 
mass in the pure quark-scalar diquark
model, leads to unphysical results in QM. Here again we face the problem that 
the parameters
which work for the nucleon and NM lead to unphysical results when used in QM.
However,  we 
believe that for the
case of $r_s$ the reason is more simple. 
The pure quark-scalar diquark model attributes 
the whole attraction
in the 3-quark system to the scalar diquark correlations, whereas 
it is known that there 
are other important
attractive effects, such as pion exchange \cite{CBM,RecentPaperwithRoberts}. 
Indeed, it has been shown in Ref.\cite{PION} 
that the inclusion of pion exchange
leads to a large reduction of $r_s$ in order to reproduce the experimental nucleon mass. 
The axial vector
diquark channel also leads to some reduction of $r_s$ \cite{MIN}. It therefore seems that 
$r_s=0.51$ is actually an overestimate, and if $r_s$ is determined by using a more complete
model for the nucleon, it might also work in the QM calculation.  

Let us now illustrate the NM $\rightarrow$ QM phase transition, taking the case 
$r_s=0.2$ of Fig. 8 as an example. Since for a given $\mu$ the actual 
ground state of the system is the state
with the largest pressure, we see from Fig. 8 
(and the insert shown in Fig. 5) that below the
NM saturation point the vacuum (VAC) is the ground state, 
and then the phase transitions to  
NM and QM occur successively. Since both phase transitions are of first order, 
the chemical potential is
continuous, but the density is discontinuous across the transitions. 
In Fig. 11 we show the
density of the ground state as a function of $\mu$. The density jump for the NM$\rightarrow$ 
QM transition
is quite large, about $0.4$ fm$^{-3}$. Fig. 12 shows the constituent 
quark mass and the gap 
of the
ground state as functions of $\mu$. This figure illustrates again that the NM state shows 
only little
tendency toward chiral symmetry restoration, and the transition 
leads to a QM state which 
is (almost) chirally symmetric, but where the color symmetry is strongly 
broken. In Figs. 13 and 14 
we show the pressure and the energy
density of the ground state as functions of the density. In the mixed phases, which are 
indicated by the dashed lines, the pressure is constant and the energy density increases
linearly with density. 

\section{Summary and conclusions}

In order to discuss the behavior of baryonic matter over a wide range of densities,
one has to describe the single hadron, normal saturating nuclear matter, and high
density quark matter at the same time. In this paper we have shown a consistent
formulation and a feasible numerical treatment of these three aspects, which makes
use of the quark-diquark picture for the single nucleon and the Hartree approximation
for the many-body systems. We used the NJL model as an effective quark theory, since this
model seems to be a unique candidate to provide a simultaneous covariant description
of all three aspects of the problem. We paid special attention to the question of those
conditions under which a phase transition from nuclear to quark matter becomes possible.

For the description of normal saturating nuclear matter, there are two important
ingredients in the NJL model. First, one has to take
into account one aspect of confinement physics, 
namely the absence of unphysical 
thresholds for the decay into quarks. In the NJL model this can be realized by
introducing an infrared cut-off in the framework of the 
proper time regularization
scheme. Only if these unphysical thresholds are avoided, can one 
describe normal
saturating nuclear matter and stable hadrons in the medium. 
Second, the vector interaction leads to a repulsion between the
nucleons which is inevitable for describing saturation. 

These two ingredients are very important in nuclear matter, but it is not clear
a priori whether they simply can be taken over to quark matter. 
In this paper we 
presented arguments which
indicate that it is preferable to use no infrared cut-off and no vector interaction in the high 
density quark phase.~\footnote{We remind the reader that, 
concerning the equation of state of 
high density quark matter, the value of the infrared cut-off has no strong influence on the 
results, but the strength of the vector interaction is crucial. 
There would be no phase 
transition from nuclear matter to quark matter if one used 
the same strength of the vector interaction 
for the description of the two phases.} First, the use 
of an infrared cut-off would lead to hadronic
poles for all densities in quark matter, 
and to a threshold behavior of the gap for small 
pairing strength
and/or small chemical potential, which contradicts  
the results obtained on more general grounds. 
It is therefore preferable to use no infrared cut-off in quark matter. 
Second, the use of the same vector interaction as in nuclear matter would lead to $\omega$ meson poles
in the high density region of quark matter.

After the discussion of these points, we turned the problem of the phase transition from
nuclear matter to normal quark matter. We found that there is no phase transition 
at all -- i.e., nuclear matter is the ground state for all 
densities. We then took into account the 
effect of scalar diquark condensation in quark matter,
treating the 4-fermi coupling constant in this channel (the pairing strength) as
a free parameter. We found that the quark matter equation of state is softened considerably,
even for relatively small pairing strengths, leading to a first order phase 
transition from nuclear matter to quark matter at a transition density which decreases
with increasing pairing strength. That is, in our calculation there is a phase transition
from nuclear matter to superconducting quark matter, but no 
phase transition to normal quark matter.
The ground state of the system is then nuclear matter at normal densities, where chiral
symmetry is strongly broken and color symmetry is intact, and quark matter 
at high densities, where chiral symmetry is almost restored and color symmetry
is strongly broken. We illustrated this picture by using an explicit numerical example
in our model calculation. 

The results discussed in this paper for the equations of state of nuclear matter and quark matter show 
that, in many respects, these two systems have very different properties. 
In order to discuss the phase transition more quantitatively, it is important to
account for these differences in the framework of an effective theory which is able to
describe both phases consistently.     

\vspace{1 cm}

{\sc Acknowledgement} \\

This work was supported by the Grant in Aid for Scientific Research of the Japanese Ministry of 
Education, Culture, Sports, Science and Technology, Project No. C2-13640298,  
the Australian Research Council and The University of Adelaide.

\newpage

\appendix
{\LARGE Appendices}
\section{Hadronization in the path integral formalism}
\setcounter{equation}{0}
In this Appendix we present some formulae which are used in sect. 3.1. The Lagrangian density is given
by (\ref{lag}), supplemented by (\ref{add}).

\subsection{Functional integration formulae}
To derive (\ref{leff1}), (\ref{leff2}), the integration over the quark fields is done by using the formula
\begin{eqnarray}
\int {\cal D}{\psi} \int {\cal D}{\overline \psi}\, {\rm exp}\left(i \left[
{\overline \Psi} S^{-1} \Psi + {\overline \Psi}\xi + {\overline \xi}\Psi \right]\right)
= {\rm exp} \left(i \left[- \frac{i}{2} {\rm Tr}\, {\rm ln}\, S^{-1} + \overline{\xi} S \xi \right]\right).
\nonumber \\
\label{fermi1}
\end{eqnarray}

To derive (\ref{dint1}), (\ref{dint2}), we replace ${\displaystyle \Delta\rightarrow \frac{\delta}{\delta\left(iJ^*\right)}}$,
${\displaystyle \Delta^*\rightarrow \frac{\delta}{\delta\left(iJ\right)}}$ in the interaction part ${\cal S}_{DI}$, and
use the formula
\begin{eqnarray}
\int {\cal D}\Delta \int {\cal D}\Delta^* {\rm exp}\left(i\left[\Delta D^{-1} \Delta^* + \Delta^*J 
+ J^* \Delta\right]\right) = {\rm exp} \left(i\left[i\,{\rm Tr\,ln}\, D^{-1} - J^* D J\right]\right).
\nonumber \\
\label{bose}
\end{eqnarray}

Finally, to derive (\ref{trlog}), one integrates over the nucleon sources ${\overline \phi}$ and $\phi$
by using a relation analogous to (\ref{fermi1}), i.e.,
\begin{eqnarray}
\int {\cal D}{\phi}\int {\cal D}{\overline \phi}\, {\rm exp}\left(i \left[
{\overline \phi} G_N \phi + {\overline \phi} N + {\overline N} \phi \right]\right)
= {\rm exp} \left(i \left[-i\,  {\rm tr}\, {\rm ln}\, G_N + \overline{N}\, G_N\, N \right]\right).
\nonumber \\
\label{nucl}
\end{eqnarray}
Note that the trace symbol Tr in (\ref{fermi1}) includes a trace in Nambu-Gorkov space, in contrast
to the trace in (\ref{nucl}), and therefore there is no factor $\frac{1}{2}$ in the exponent of
(\ref{nucl}).

\subsection{Derivation of Eq.(\ref{term})}
The first term in (\ref{leff2}) involves the propagator $S$, which is obtained
by inverting (\ref{ipr}). If we write
\begin{eqnarray}
S = \left(
\begin{array}{cc}
S_{11} & S_{12} \\
S_{21} & S_{22}
\end{array} \right) \label{s}
\end{eqnarray}
then the diagonal elements are obtained as
\begin{eqnarray}
S_{11}=\frac{S_0}{1+G_0 \gamma_5 \Delta_a \beta_a \tilde{S}_0 \gamma_5 \Delta^*_a \beta_a},
\,\,\,\,\,\,\,\,\,\,\,\,\,\,\,
S_{22}=\frac{{\tilde S}_0}{1+{\tilde S}_0 \gamma_5 \Delta^*_a \beta_a S_0 \gamma_5 \Delta_a \beta_a}\,\,,
\nonumber
\end{eqnarray}
while the non-diagonal elements are
\begin{eqnarray}
S_{12} = i \,S_{11} \gamma_5 \Delta_a \beta_a {\tilde S}_0,
\,\,\,\,\,\,\,\,\,\,\,\,\,\,\,
S_{21} = i \,S_{22} \gamma_5 \Delta^*_a \beta_a S_0. \nonumber 
\end{eqnarray}
The quantities $S_0$ and ${\tilde S}_0$ are given in momentum space by (\ref{s0}).
When the above form for $S$ is inserted into (\ref{leff2}), the non-diagonal elements of $S$ lead to
terms which involve different numbers of $\Delta$ or $\Delta^*$, and these terms vanish after the 
integration over the diquark fields. Therefore the first term in (\ref{leff2}) effectively becomes
\begin{eqnarray}
\frac{1}{3} {\overline \Phi} \left(
\begin{array}{cc}
\Delta S_{11} \Delta^* & 0 \\
0 & \Delta^* S_{22} \Delta 
\end{array} \right) \Phi.
\end{eqnarray}
Then one uses the Nambu-Gorkov forms of ${\overline \Phi}$ 
and $\Phi$, which are
expressed similar to Eq.(\ref{fields}) in terms 
of ${\overline \phi}$ and $\phi$, and the
relations $C {\tilde S}_0^T C^{-1} = S_0$, $C S_0^T C^{-1} ={\tilde S}_0$, 
where
the symbol $T$ (transpose) stands also w.r.t. the functional 
matrix structure, i.e., it implies
$x \leftrightarrow x'$. In this way one arrives at Eq.(\ref{term}).

\subsection{Derivation of Eq. (\ref{nprop})}
Let us extract the term $\propto {\overline \phi} \phi$ from (\ref{dint1})  
and (\ref{dint2}).
The second exponential factor in (\ref{dint1}) gives
\begin{eqnarray}
\lefteqn{{\rm exp}\left(- {\rm Tr}\,{\rm ln}
\left[1+\frac{1}{3}D_0 \overline{\phi}
S_0  \phi \right]\right)} \label{gn1}  \\
& & = 1- {\overline \phi} \left(S_0 D_0\right)\phi + \dots 
\equiv 1 + i \,{\overline \phi}\, \Pi_N \, \phi + \dots, \label{gn2}
\end{eqnarray}
where a color factor $3$ comes from the color trace in (\ref{gn1}), 
and $\Pi_N\equiv i S_0 D_0$
as in (\ref{bubn}).
Next, the factor in (\ref{dint2}) gives to ${\cal O}
\left({\overline \phi}\phi\right)$
\begin{eqnarray}
\lefteqn{\left[{\rm exp} \left(i{\cal S}_{DI} 
\left({\hat \Delta},{\hat \Delta}^* \right) \right) 
\,\,{\rm exp}\left(-i J^* \,D\, J\right)
\right]_{J=J^*=0}} \nonumber \\ 
& & = 1 + i\,\left[{\cal S}_{DI}\left({\hat \Delta},{\hat \Delta}^*\right) 
{\rm exp}\left(-i J^* \,D_0\, J\right)
\right]_{J=J^*=0} + \dots \label{gn3} \\
& & = 1 + \frac{i}{3} {\overline \phi} \left\{ \left[{\hat \Delta}\,S_0 
\frac{1}{M} \left({\hat \Delta}_{a'} \beta_{a'}\right)
\left({\hat \Delta}^*_{a} \beta_{a}\right) S_0 {\hat \Delta}^* \right. 
\right. \nonumber \\
& & \left. \left. + {\hat \Delta} S_0 \frac{1}{M} \left({\hat \Delta}_{a'} 
\beta_{a'}\right)
\left({\hat \Delta}^*_{a} \beta_{a}\right) \frac{1}{M} S_0 
\left({\hat \Delta}_{a'} \beta_{a'}\right)
\left({\hat \Delta}^*_{a} \beta_{a}\right) S_0 {\hat \Delta}^* + 
\dots \right] \phi \right. \nonumber \\ 
& & \left. \times {\rm exp}\left(-i J^* \,D_0\, J \right) \right\}_{J=J^*=0} + \dots, \label{gn4}
\end{eqnarray}     
where ${\displaystyle {\hat \Delta}\equiv \frac{\delta}{\delta(iJ^*)}}$, 
${\displaystyle {\hat \Delta}^* \equiv \frac{\delta}{\delta(iJ)}}$.
A contraction is obtained if one ${\hat \Delta}$ and one ${\hat \Delta}^*$ hit the same factor
$(-i J^* D_0 J) = (iJ^*)(i D_0) (iJ)$, which appears in the expansion of the exponent in (\ref{gn4}).
This contraction of ${\hat \Delta}_a$ with ${\hat \Delta}_{a'}$ gives a factor $i D_0 \delta_{a'a}$.  
Among the many contractions which emerge in (\ref{gn4}), the ladder graphs appear as a subset besides self
interaction graphs \cite{FAD1}. To obtain the ladder graphs from (\ref{gn4}), one contracts  
${\hat \Delta}$ on the most left side with ${\hat \Delta}$ to the nearest to the right of it, then the next    
${\hat \Delta}$ on the left side with ${\hat \Delta}$ nearest to the right of it, etc.
To write down the resulting series, we make the color structure more explicit by decomposing   
\begin{eqnarray}    
\left[\left({\hat \Delta}_{a'} \beta_{a'}\right) \left({\hat \Delta}^*_{a} \beta_{a}\right)\right]_{kl}
\equiv C^{ab}_{kl} {\hat \Delta}^*_a {\hat \Delta}^*_b 
= \left(-3\,P^{(0)ab}_{kl} + \frac{3}{2} P^{(8)ab}_{kl}\right) {\hat \Delta}^*_a {\hat \Delta}^*_b\,\,, \nonumber 
\end{eqnarray}
where the color singlet and octet projection operators are given by:
\begin{eqnarray}
P^{(0)ab}_{kl} = \frac{1}{3} \delta_{ka}\delta_{lb} \,\,\,\,\,\,\,\,\,\,\,\,\,\,\,
P^{(8)ab}_{kl} = \delta_{kl}\delta_{ab}-\frac{1}{3} \delta_{ka}\delta_{lb} \nonumber 
\end{eqnarray}
Then the ladder-type contribution from (\ref{gn4}) is given by
\begin{eqnarray}
1 &+& \frac{i}{3} {\overline \phi} \left[\Pi_N \,\frac{1}{M}\, \Pi_N\, C^{ij}_{ij} +
\Pi_N \,\frac{1}{M}\, \Pi_N \,\frac{1}{M}\, \Pi_N\, C^{ib}_{ib}C^{bj}_{lj} + \dots \right]\phi \nonumber \\
&=& 1 + i {\overline \phi} \,\Pi_N\, T_N\, \Pi_N \phi \label{gn5}
\end{eqnarray}
where the quark-diquark T-matrix is given in (\ref{tn}). The sum of 
(\ref{gn2}) and (\ref{gn5}) is then
$1+i{\overline \phi} G_N \phi = {\rm exp}\left(i\,{\overline \phi} 
G_N \phi\right)+\dots$ , 
where the quark-diquark propagator in the color singlet channel, $G_N$, 
is given by (\ref{gn}). In this way one arrives at (\ref{nprop}).

\section{Explicit evaluation of the Dirac determinant}
\setcounter{equation}{0}
The contribution of the quark loop to the effective action is 
(see Eq.(\ref{qdet}))  
${\displaystyle -\frac{i}{2}{\rm Tr}\,{\rm ln}\,S^{-1}} \equiv - \int {\rm d}^4 x \,V_{\ell}$, 
where the inverse quark
propagator is a $2 \times 2$ matrix in Nambu-Gorkov space and given by (\ref{iprq}). 
By using the relation
\begin{eqnarray}
{\rm Tr}\,{\rm ln} \left(
\begin{array}{cc}
A & B \\ 
C & D 
\end{array} \right) = {\rm Tr}\,{\rm ln}\left(-B\,C + B\,D\,B^{-1}\,A\right)
\nonumber
\end{eqnarray}
and ${\displaystyle \beta_1^2=\frac{3}{2} {\rm diag}\left(0,1,1\right)
\equiv \frac{3}{2} C_1}$, 
we obtain 
\begin{equation}
V_{\ell}= \frac{i}{2} \int \frac{{\rm d}^4p}{\left(2\pi\right)^4} {\rm Tr}\,{\rm ln}
\left(\Delta^2\,C_1+\left(-\fslash{p}-M+\nu \gamma_0\right)
\left(\fslash{p}-M+\nu\gamma_0\right)\right), \label{vl}
\end{equation}
where we defined $\nu\equiv \mu_q^*$ to simplify the notation.
For the evaluation of the Dirac determinant it is sufficient, 
because of rotational invariance, 
to consider the vector ${\bold p}$ along the z-axis, and (\ref{vl}) becomes
\begin{eqnarray}
V_{\ell}= \frac{i}{2} \int \frac{{\rm d}^4p}{\left(2\pi\right)^4} 
{\rm Tr}\,{\rm ln}
\left(\Delta^2\,C_1 -p_0^2+p^2+M^2+\nu^2-2M\nu \gamma^0 + 2\nu p 
\gamma^3 \gamma^0\right),
\nonumber
\end{eqnarray}
where in this Appendix we use the notation 
$p^2\equiv {\bold p}^2$ and $E_p\equiv \sqrt{p^2+M^2}$.
By using explicit representations for the Dirac matrices, it is easy to 
calculate the
Dirac determinant explicitly. Because of isospin we get a factor 2, 
and one is left only
with the color determinant:
\begin{eqnarray}
V_{\ell}= 2i \int \frac{{\rm d}^4p}{\left(2\pi\right)^4} {\rm ln}\,{\rm det}_C
\left[\left(\Delta^2\,C_1 -p_0^2+E_p^2+\nu^2\right)^2-4 E_p^2 \, \nu^2\right]. 
\end{eqnarray}
By using $C_1={\rm diag}\left(0,1,1\right)$, 
one can calculate the color determinant
with the result
\begin{eqnarray}
V_{\ell}= 2i \int \frac{{\rm d}^4p}{\left(2\pi\right)^4} \left\{
2\,{\rm ln}\left[\left(p_0^2-\epsilon_+^2\right)
\left(p_0^2-\epsilon_-^2\right)\right]
+{\rm ln}\left[\left(p_0^2-\epsilon_{0+}^2\right)
\left(p_0^2-\epsilon_{0-}^2\right)\right] \right\}, \nonumber \\
\end{eqnarray}
where $\epsilon_{\pm}^2=(E_p\pm\nu)^2+\Delta^2$ and 
$\epsilon_{0\pm}^2=(E_p\pm\nu)^2$. In this way one arrives at (\ref{qdet}). 

It is convenient to separate the part which survives for $\Delta=0$ 
from the rest. The latter contribution to the effective potential is given by the first line
of (\ref{vd}). The former contribution becomes, after subtracting the zero density value,
\begin{eqnarray}
V_{\ell}(\Delta=0)= 6i \int \frac{{\rm d}^4p}{\left(2\pi\right)^4} \left[
{\rm ln}\,\frac{p_0^2-(E_p+\nu)^2}{p_0^2-E_{0p}^2}
+{\rm ln}\,\frac{p_0^2-(E_p-\nu)^2}{p_0^2-E_{0p}^2}\right],
\end{eqnarray}
where $E_{0p}=\sqrt{M_0^2+p^2}$.
This can further be split into the pure vacuum loop which contributes to $V_{\rm vac}$
given in (\ref{vac}), and the density dependent term
\begin{eqnarray}
V_Q &=& 6i \int \frac{{\rm d}^4p}{\left(2\pi\right)^4} \left[
{\rm ln}\,\frac{p_0^2-(E_p+\nu)^2}{p_0^2-E_{p}^2}
+{\rm ln}\,\frac{p_0^2-(E_p-\nu)^2}{p_0^2-E_{p}^2}\right] \nonumber \\
&=& -6 \int \frac{{\rm d}^3p}{\left(2\pi\right)^3} \left[
\left(E_p+\nu\right)-E_p + |E_p-\nu| -E_p\right] \nonumber \\
&=& -12  \int \frac{{\rm d}^3p}{\left(2\pi\right)^3} \Theta(\nu-E_p)\left(\nu-E_p\right),
\end{eqnarray}
which is Eq.(\ref{fermiq}). 

We also note that the evaluation of the nucleonic 
contribution, (\ref{vn}), to the effective 
potential for NM proceeds similarly, by setting $\Delta=0$ in (\ref{vl}) 
and replacing
$M\rightarrow M_N$, $\nu\rightarrow \mu^*$. (Of course, there is no color factor 3 in this
case.) Since the contribution for $\mu=0$ is subtracted, one is left with the density
dependent piece (\ref{fermi}).

\section{The $\omega$ meson in nuclear and quark matter}
\setcounter{equation}{0}
The expression for the the $q{\overline q}$ bubble graph in quark matter
(Fermi momentum $p_F$) for an external isoscalar vector field with $q^{\mu}=(q^0,{\bold 0})$
is given by
\begin{eqnarray}
\Pi^{\mu \nu}(q)=\int \frac{{\rm d}^4k}{\left(2\pi\right)^4} \Theta(|{\bold k}|-p_F)
{\rm Tr} \frac{\gamma^{\mu}(\fslash{k}+M)\gamma^{\nu}(\fslash{k}+\fslash{q}+M)}
{\left(k^2-M^2\right)\left(\left(k+q\right)^2-M^2\right)}. \label{piv}
\end{eqnarray}
We introduce a Feynman parameter ($x$) and make a shift $k\rightarrow k-qx$. Since 
${\bold q}=0$, this shift affects only the time component, which is no problem since the
proper time regularization scheme does not directly cut the loop momentum.
The evaluation of the Dirac trace gives a term $\propto k^{\mu}k^{\nu}$, for
which we perform a partial integration, thereby picking up a surface term arising from the
$\Theta$ function in (\ref{piv}). The result is $\Pi^{00}(q)=0$ and 
$\Pi^{ij}(q)\equiv g^{ij} \Pi(q)$ with
\begin{eqnarray}
\Pi(q)&=&-48\, q_0^2 \int \frac{{\rm d}^4k}{\left(2\pi\right)^4} \Theta(|{\bold k}|-p_F)
\int_0^1{\rm d}x \frac{x(1-x)}{\left(k^2+M^2-q_0^2x(1-x)\right)^2} \nonumber \\ \label{piva}  \\
&+& 8 p_F \int \frac{{\rm d}^4k}{\left(2\pi\right)^4} \delta(|{\bold k}|-p_F)
\int_0^1{\rm d}x \frac{1}{k^2+M^2-q_0^2x(1-x)} \label{pivb}
\end{eqnarray}
Here we performed a Wick rotation in $k_0$. This expression is now regularized according
to the prescription (\ref{pt}). (In order that the imaginary part is canceled
for the case $\Lambda_{\rm IR}>0$, the finite surface term (\ref{pivb}) has to
be treated in the same way as the divergent part (\ref{piva}).) For the part (\ref{piva}) we 
introduce Euclidean polar coordinates
($k,\Theta)$ by $|{\bold k}|=k \sin \Theta$, $k_0=k \cos \Theta$, where $\Theta$ is restricted
by ${\displaystyle 0\leq \Theta \leq {\rm arc\,\, sin} \frac{p_F}{k}}$. After performing the $\Theta$ integration
we obtain for the piece (\ref{piva}), which we call ${\hat \Pi}$,
\begin{eqnarray}
{\hat \Pi}(q)&=&-\frac{3}{\pi^2} \,q_0^2\, \int_{p_F^2}^{\infty} t\,{\rm d}t\, 
G\left(\sqrt{\frac{p_F^2}{t}}\right)
\int_0^1 {\rm d}x\,x(1-x) \int_{1/\Lambda_{\rm UV}^2}^{1/\Lambda_{\rm IR}^2}
{\rm d}\tau\,\tau \nonumber \\
&\times& {\rm exp}\left(-\tau \left[t+M^2-q_0^2\,x(1-x)\right]\right), \label{piv1}
\end{eqnarray}
where 
\begin{eqnarray}
G(y)=1-\frac{2}{\pi}\left({\rm arc\,\,sin} y -y\sqrt{1-y^2}\right). \nonumber
\end{eqnarray} 
The $\tau$ integration can also be performed, and the final result is
\begin{eqnarray}
{\hat \Pi}(q)&=&-\frac{3}{\pi^2}\, q_0^2\, \int_{p_F^2}^{\infty} t\,{\rm d}t\, 
G\left(\sqrt{\frac{p_F^2}{t}}\right)
\int_0^1 {\rm d}x\,x(1-x) \frac{1}{A^2} \nonumber \\
&\times& \left[e^{-A/\Lambda_{\rm UV}^2} \left(1+\frac{A}{\Lambda_{\rm UV}^2}\right) 
- e^{-A/\Lambda_{\rm IR}^2} \left(1+\frac{A}{\Lambda_{\rm IR}^2}\right) \right], \label{fin}
\end{eqnarray}
where $A=t+M^2-q_0^2\,x(1-x)$. For $\Lambda_{\rm IR}>0$, $\left[\dots\right]$ in (\ref{fin})
behaves $\propto A^2$ for $A\rightarrow 0$, and there is no imaginary part.
The piece (\ref{pivb}) is given by
\begin{eqnarray}
\delta \Pi(q)= \frac{4p_F^3}{\pi^3} \int_0^{\infty}{\rm d}k_0 \int_0^1{\rm d}x
\frac{{\rm exp}\left(-B/\Lambda_{\rm UV}^2 \right)-
{\rm exp}\left(-B/\Lambda_{\rm IR}^2 \right)}{B}, \nonumber
\end{eqnarray}    
where $B=k_0^2+M^2+p_F^2-q_0^2\,x(1-x)$.

\section{Numerical results for other choices of parameters}
\setcounter{equation}{0}
In this Appendix we present some numerical results for the QM EOS 
obtained by using 
values of $\Lambda_{\rm IR}$ or $r_{\omega}$ other than those used 
in the main text. 

First we consider the case where the same strength of the vector interaction as in NM ($r_{\omega}=0.37$) is
used also in QM, taking $\Lambda_{\rm IR}=0$ in QM. The results are shown in Fig. 15.
The dotted line here ($r_s=0$ in QM) is very similar to the dashed line for $r_{\omega}=0.37$ in Fig. 5. (We note that
in Fig. 5 we used $\Lambda_{\rm IR}=0.2$ GeV.) It is clear from Fig. 15
that it is not possible to have a NM $\rightarrow$ QM phase transition, even if one uses
large values of $r_s$.

Next we consider the case where the same value of the IR cut-off as in NM ($\Lambda_{\rm IR}=0.2$ GeV) is used also
in QM, taking $r_{\omega}=0$ in QM. The results are shown in Fig. 16.  
The dotted line here ($r_s=0$ in QM) agrees with the dashed 
line for $r_{\omega}=0$ in Fig. 5. We see that the qualitative 
behavior of the curves with increasing $r_s$ is similar to the 
case $\Lambda_{\rm IR}=0$ shown in Fig. 8,
although the sensitivity to the pairing strength $r_s$ is somewhat 
stronger for the case $\Lambda_{\rm IR}=0$. 
(For given $r_s$, the gaps
are larger for the case of $\Lambda_{\rm IR}=0$.) The case of $r_s=0.25$,  
shown in Fig. 16, is actually very
similar to the case $r_s=0.2$ in Fig. 8, which was used in sect. 4.3 
as an example to illustrate the NM $\rightarrow$ QM
phase transition.

\newpage

\section*{Tables}

\begin{table}[hbt]
\begin{center}
\begin{tabular}{|c||c|c|}
\hline
                        &  Nucleons and NM & QM \\  \hline
m [MeV]                 &   16.93          & 17.08  \\
$G_{\pi}$[GeV$^{-2}]$   &  19.60           & 19.76 \\
$\Lambda_{\rm UV}$[MeV] &  638.5           & 636.7 \\
$\Lambda_{\rm IR}$[MeV] &  200.0           & 0 \\
$r_{\omega}$            &  0.37           & 0 \\ 
$r_s$                   &  0.51            & free parameter \\  \hline  
\end{tabular}
\end{center}
\caption{Parameters used for single nucleons, nuclear matter (left column) 
and  quark matter (right column). The proper time 
regularization scheme is used in both cases.}  
\end{table}

\newpage

\section*{Figure Captions}

\begin{enumerate}
\item Nucleon mass as function of scalar potential for 
$\Lambda_{\rm IR}=0.2$ GeV and $\Lambda_{\rm IR}=0$. The parameter set for $\Lambda_{\rm IR}=0.2$ GeV
is shown in Table 1, and that for $\Lambda_{\rm IR}=0$ can be found in Table 1 of Ref.\cite{BT}. 
The quark-diquark threshold for the case $\Lambda_{\rm IR}=0$
is also shown.

\item  Binding energy per nucleon as function of density for 
$\Lambda_{\rm IR}=0.2$ GeV and $\Lambda_{\rm IR}=0$. The parameter set for $\Lambda_{\rm IR}=0.2$ GeV
is shown in Table 1, while that for $\Lambda_{\rm IR}=0$ can be found in Table 1 of Ref.\cite{BT}.

\item The nucleon and quark effective masses in NM (solid lines), and the quark
effective mass in normal (non-superconducting) QM (dashed line) are shown as functions of the density.
The parameters used in the NM calculation are shown in the second row of Table 1, and the result
for QM shown here refers to the same parameter set. 
(Note that the parameters $r_{\omega}$ and
$r_s$ are relevant only for the NM result shown in this figure.)

\item Effective mass of the $\omega$ meson at rest (${\bold q}=0$) in NM as function
of the density. The dashed line shows the density dependence of $2M^*$. The parameters used in the 
calculation are shown in the second row of Table 1.

\item $P-\mu$ plots for NM (solid line and insert) and normal (non-superconducting) QM. The two 
dashed lines show the QM result for two values of $r_{\omega}$, and the dotted lines show the massless
quark gas EOS for two values of the bag constant $B$. The parameters used in the NM calculation are 
shown in the second row of Table 1, and the results for QM 
shown here refer to the same parameter set, apart from  
the variation of the parameter $r_{\omega}$. (Both dashed lines in this figure start at the point
$(\mu,P)=(1200\, {\rm MeV}, 0)$, corresponding to zero density. For the case $r_{\omega}=0.37$ the pressure 
increases monotonously, while for $r_{\omega}=0$ the pressure first increases slightly for very small densities, 
then decreases to negative values (unstable branch), and finally increases.)

\item Effective mass of the $\omega$ meson at rest (${\bold q}=0$) in QM, 
as function
of the density, for the case $\Lambda_{\rm IR}=0.2$ GeV. 
The variation of the effective quark
mass with density used in this calculation is shown by the dashed 
line in Fig. 3. The values of 
$r_{\omega}$ used in the calculation of the effective $\omega$-meson mass 
are indicated for each line. 
The dashed line shows the density dependence of $2 \sqrt{M^{*2}+p_F^2}$.

\item Same as Fig. 6 for the case $\Lambda_{\rm IR}=0$. For 
small $r_{\omega}$, solutions exist only in the high density region.

\item $P-\mu$ plots for NM (solid line) and QM for several values of the 
pairing strength
$r_s$ in QM. The parameters used in the calculation are shown in the 
second and third rows of Table 1, 
and the values of $r_s$ in QM are indicated in the figure. A 
crossing of the NM and QM lines indicates 
a first order phase transition. The dotted line ($r_s=0$) differs slightly 
from the result shown in Figure 5 for $r_{\omega}=0$, since for Fig. 5 
the parameters shown in the 
second row of Table 1 were also used in the QM calculation.

\item The effective quark masses in NM (solid line) and QM for several 
values of
the pairing strength $r_s$ in QM as functions of the density. 
The cases 1 to 6 indicated in this figure
refer to different choices of $r_s$ in QM as indicated in Fig. 8. 
The dotted line ($r_s=0$) differs
from the result shown in Fig. 3, since for Fig. 3 
the parameters shown in the second row
of Table 1 were also used in the QM calculation.

\item The gap in QM for several values of
the pairing strength, $r_s$, as a function of the density. 
The cases 1 to 6 indicated in this figure
refer to the different choices of $r_s$ shown in Fig. 8.

\item The density of the ground state as function of the chemical potential. 
The ground state
of the system is either the vacuum (VAC), nuclear matter (NM) or 
superconducting
quark matter (QM). The pairing strength $r_s=0.2$ is used for QM.

\item The quark effective masses (solid lines) and the gap (dashed line) 
of the ground state
as functions of the chemical potential. The ground state
of the system is either the vacuum (VAC), nuclear matter (NM) or 
superconducting
quark matter (QM). The value of the gap is equal to zero for VAC and NM. 
The pairing strength $r_s=0.2$ is used for QM. 

\item The pressure in the ground state as function of the density.
The ground state of the system is either the vacuum (VAC), 
nuclear matter (NM) or superconducting
quark matter (QM). (The vacuum corresponds to the point $P=0$, $\rho=0$.) 
The dashed lines indicate 
the mixed phases. The pairing strength $r_s=0.2$ is used for QM.  

\item The energy density of the ground state as function of the density.
The ground state of the system is either the vacuum (VAC), 
nuclear matter (NM) or superconducting
quark matter (QM). (The vacuum corresponds to the 
point ${\cal E}=0$, $\rho=0$.) The dashed lines indicate 
the mixed phases. The pairing strength $r_s=0.2$ is used for QM.  

\item $P-\mu$ plots for NM (solid line) and QM for 
several values of the pairing strength, 
$r_s$, in QM. The case shown here refers to $\Lambda_{\rm IR}=0$ 
and $r_{\omega}=0.37$ in QM.

\item $P-\mu$ plots for NM (solid line) and QM for several 
values of the pairing strength, 
$r_s$, in QM. The case shown here refers to $\Lambda_{\rm IR}=0.2$ GeV 
and $r_{\omega}=0$ in QM.

\end{enumerate}

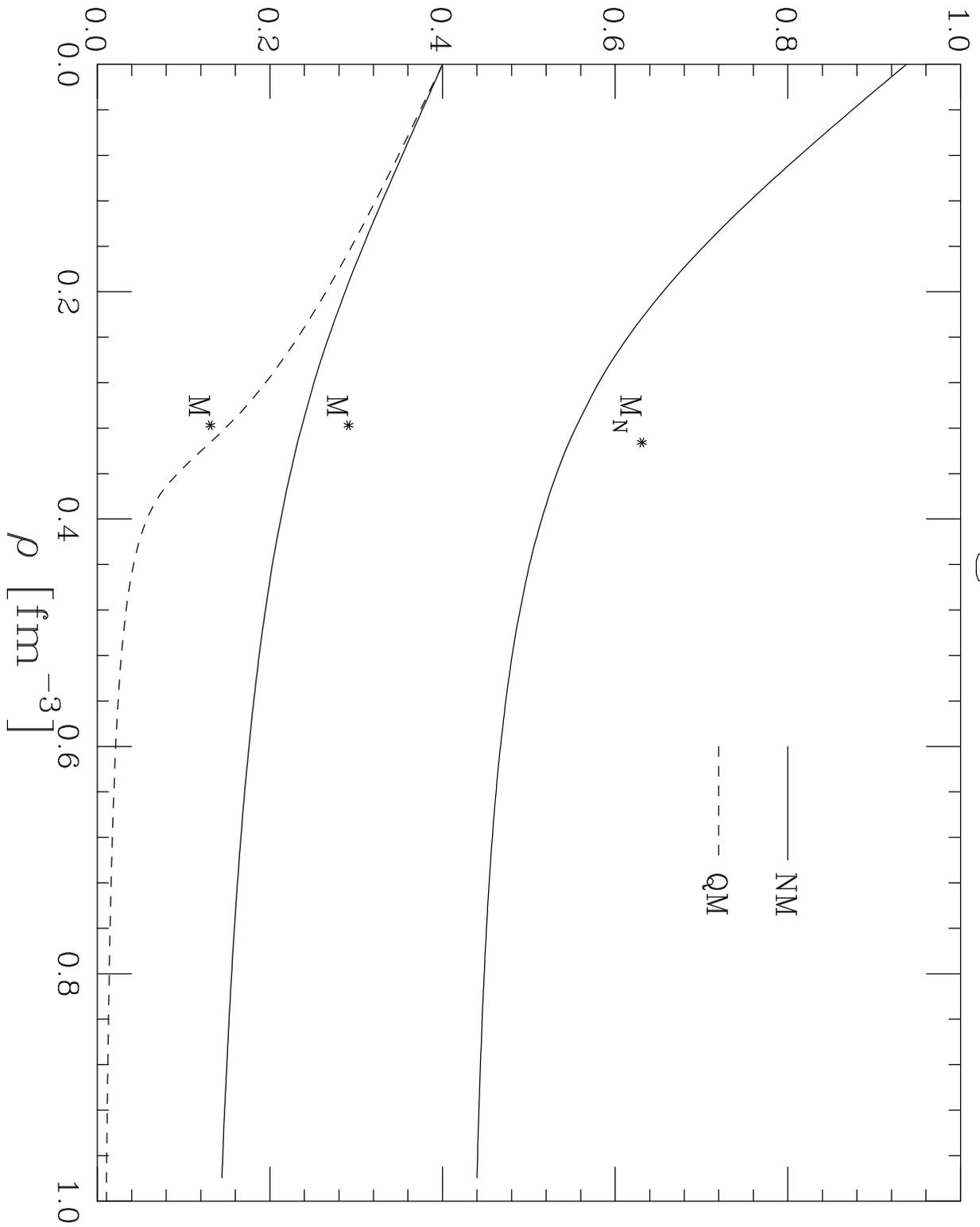
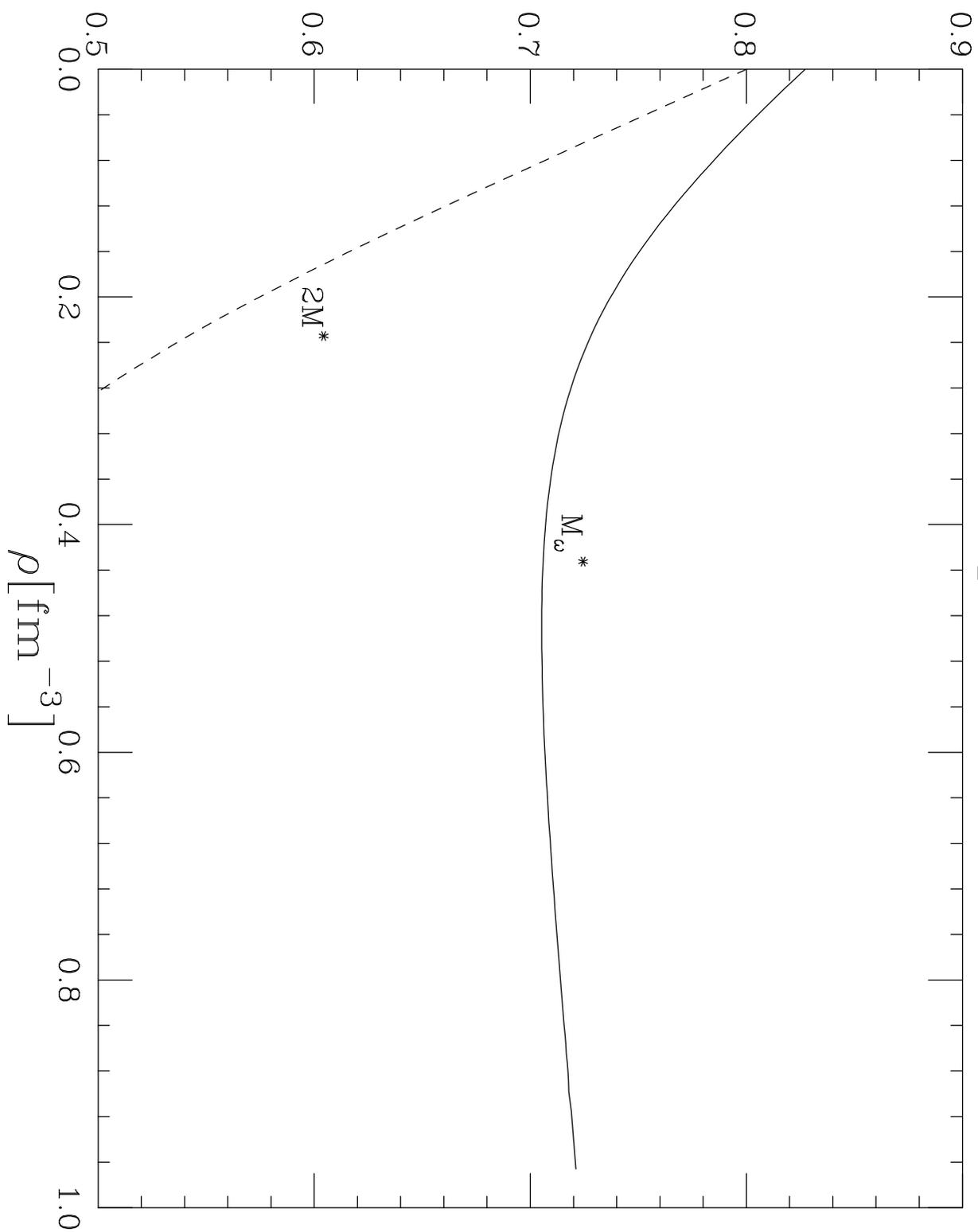
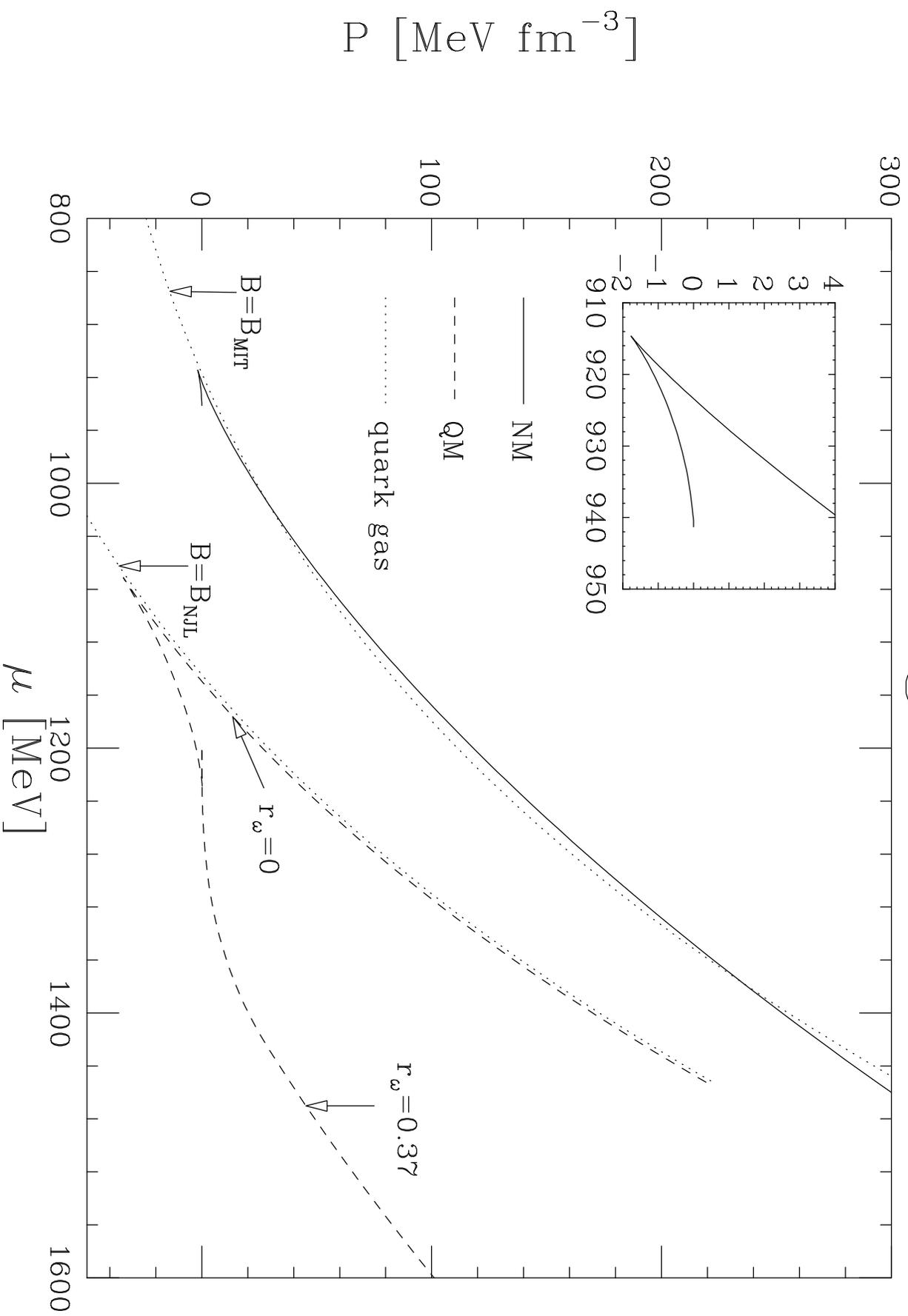
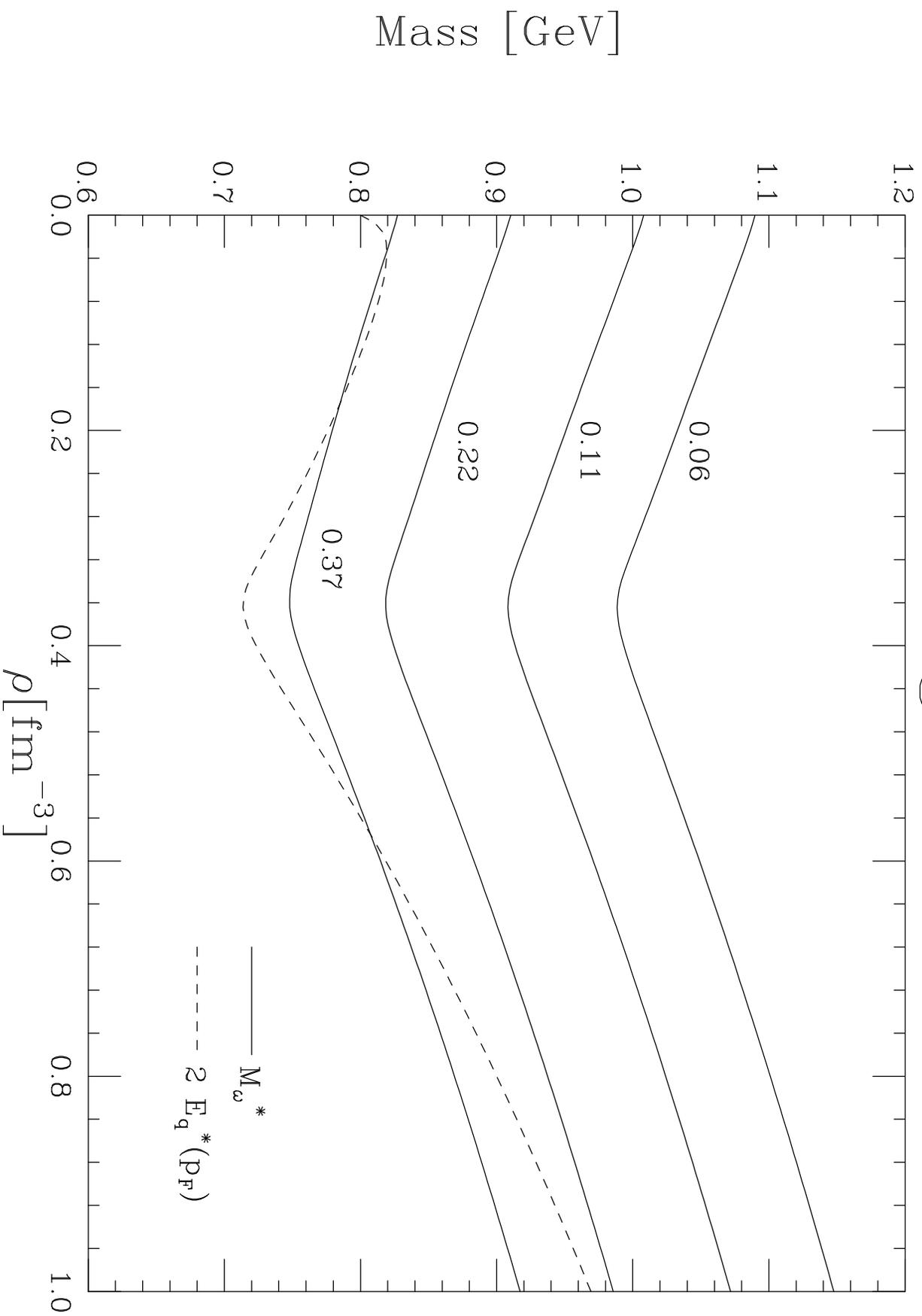
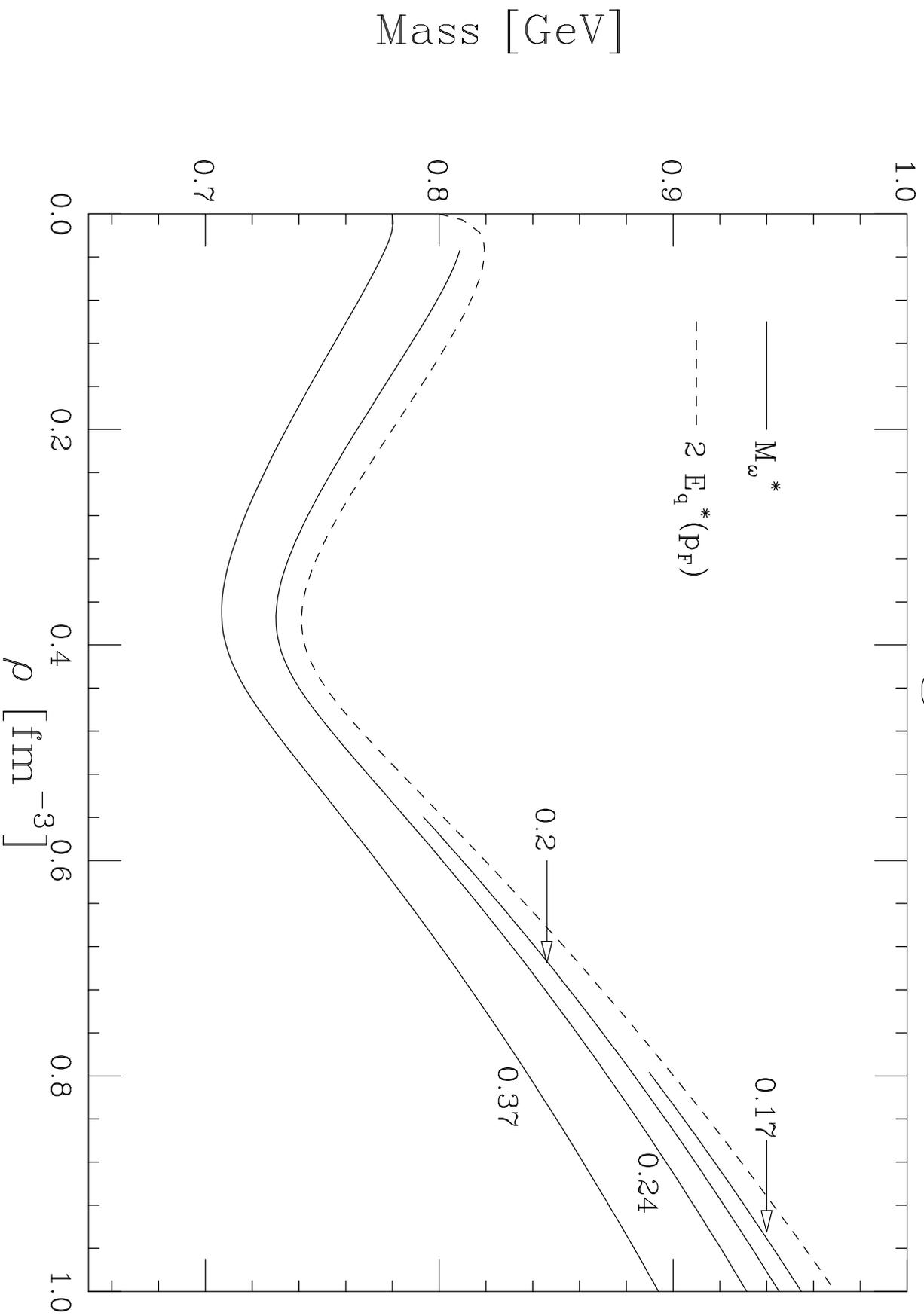
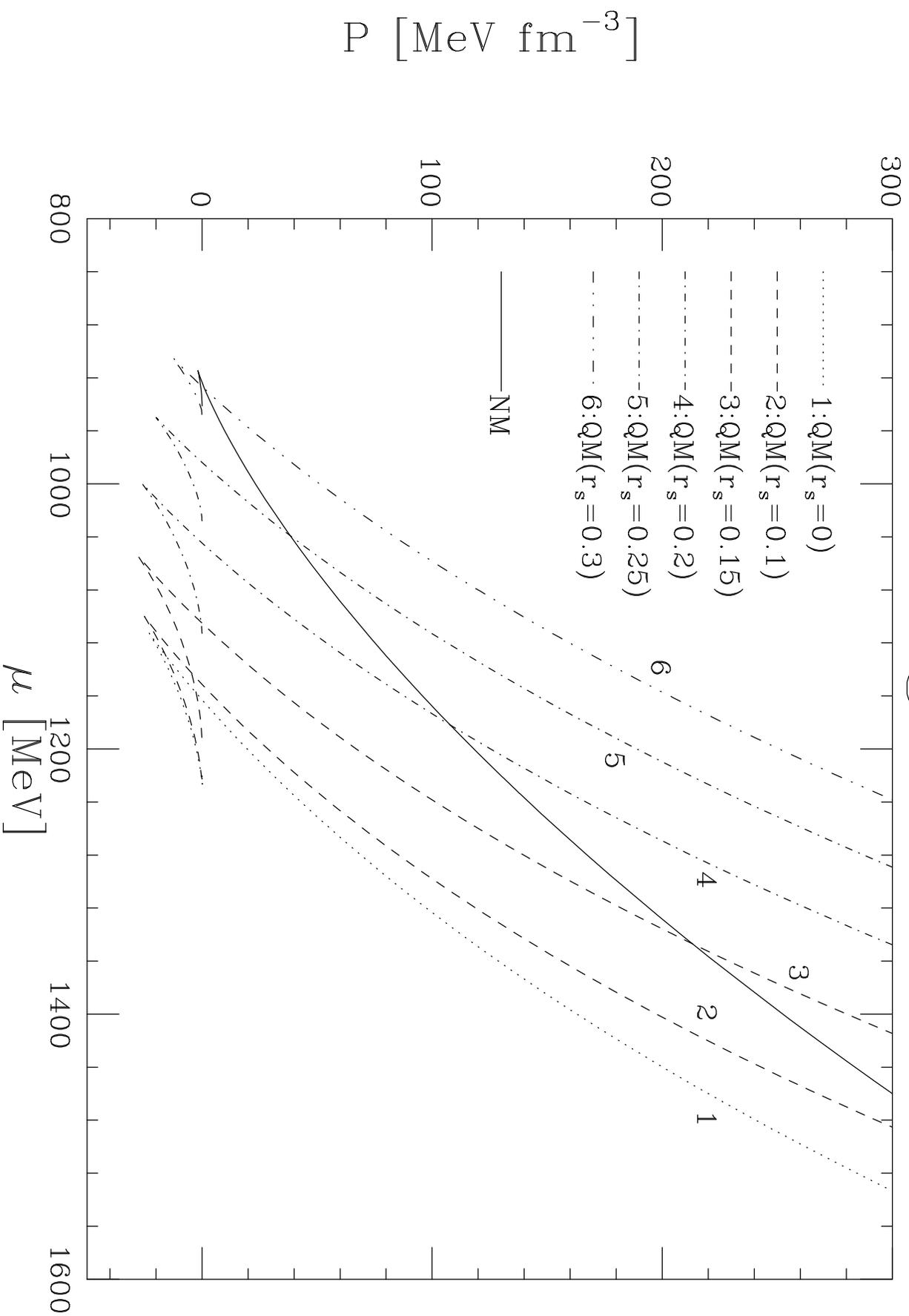
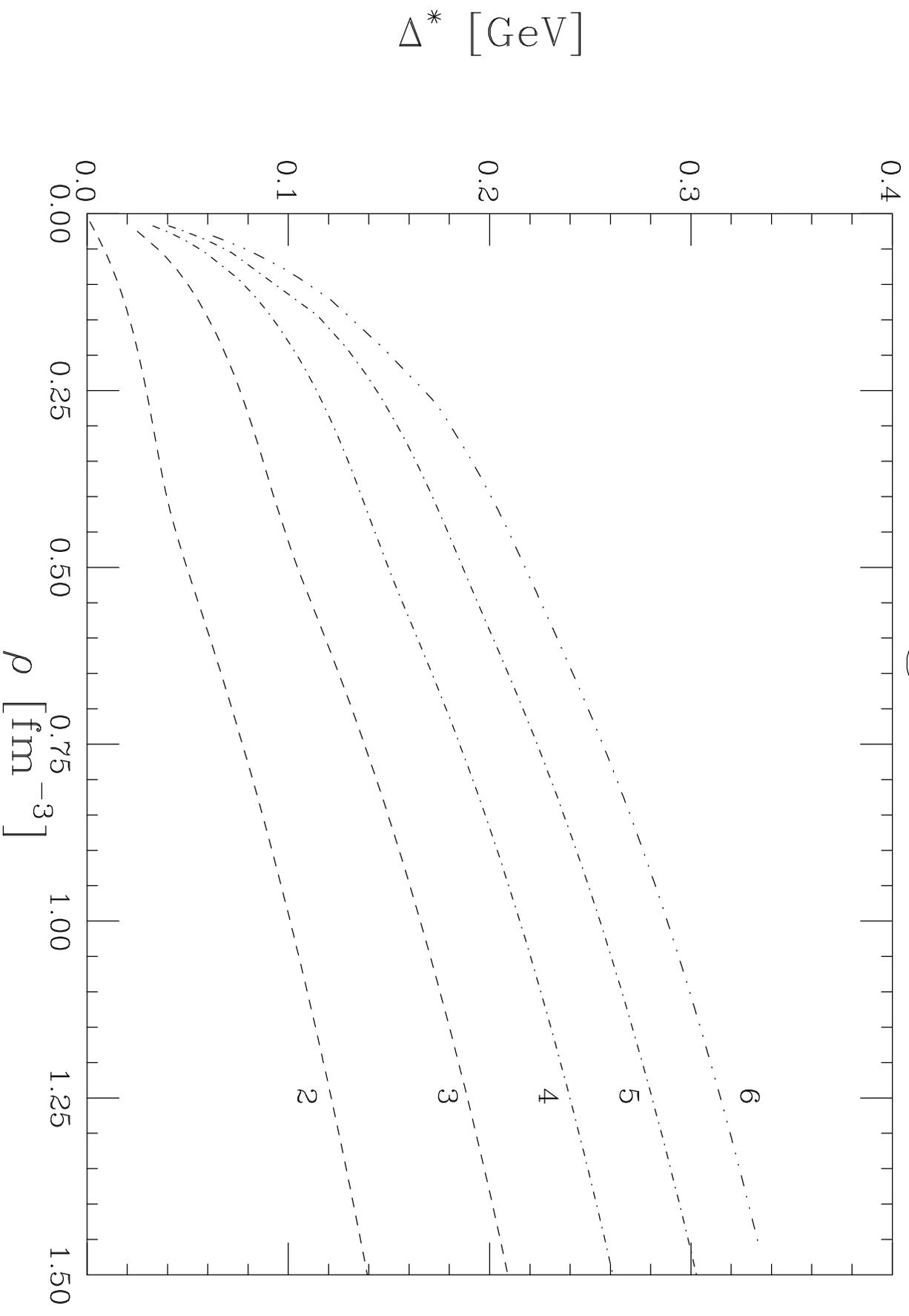
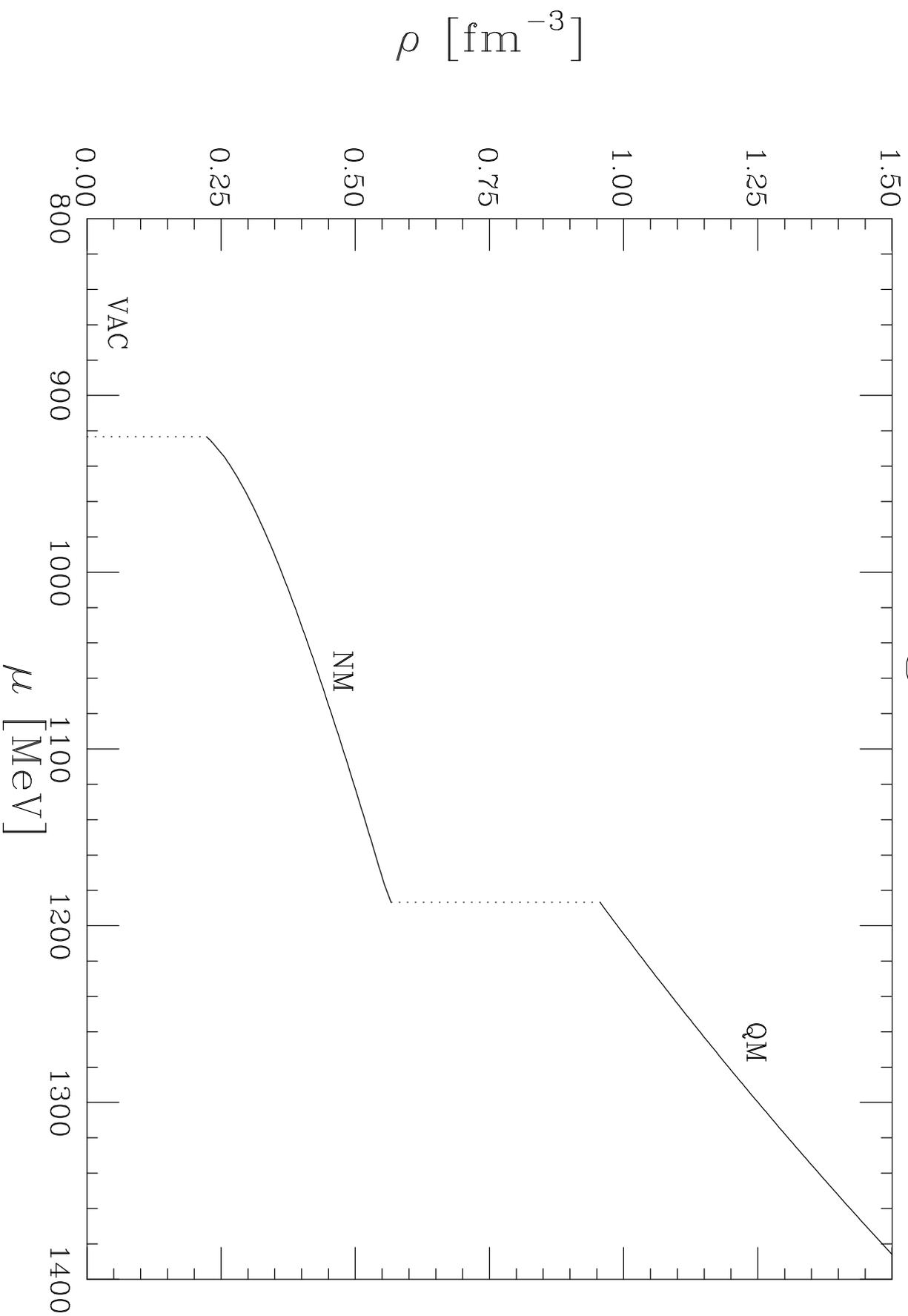
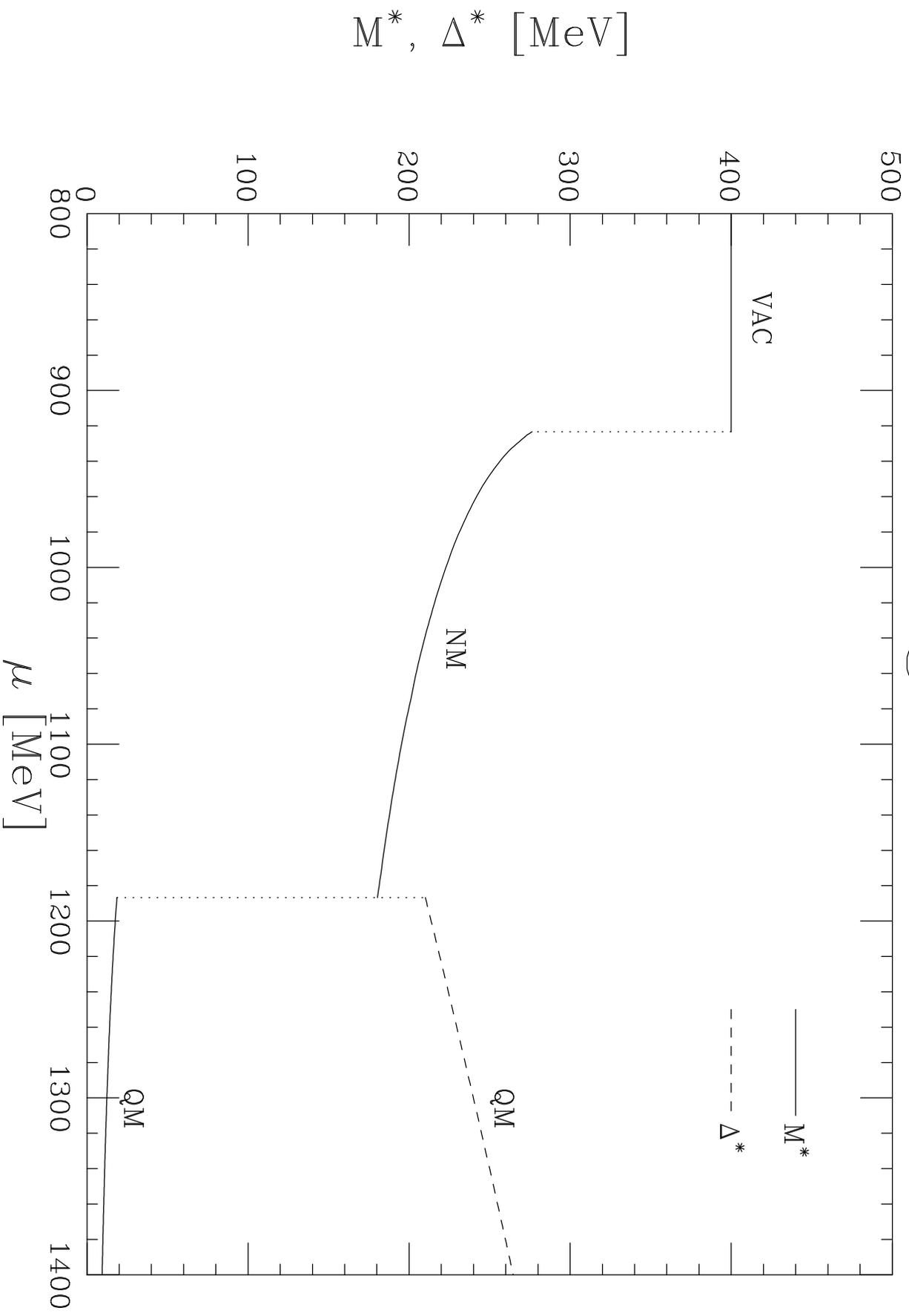
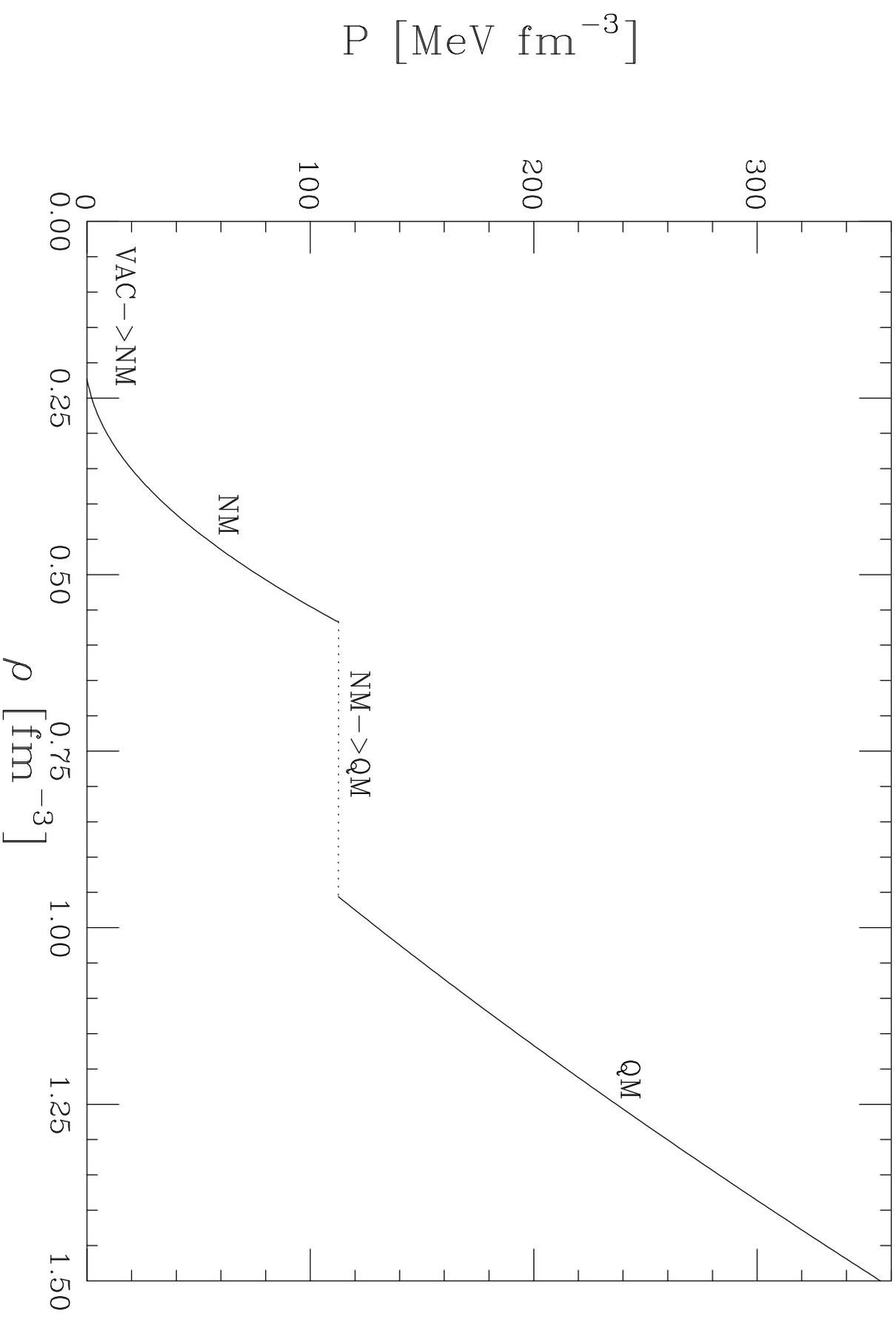
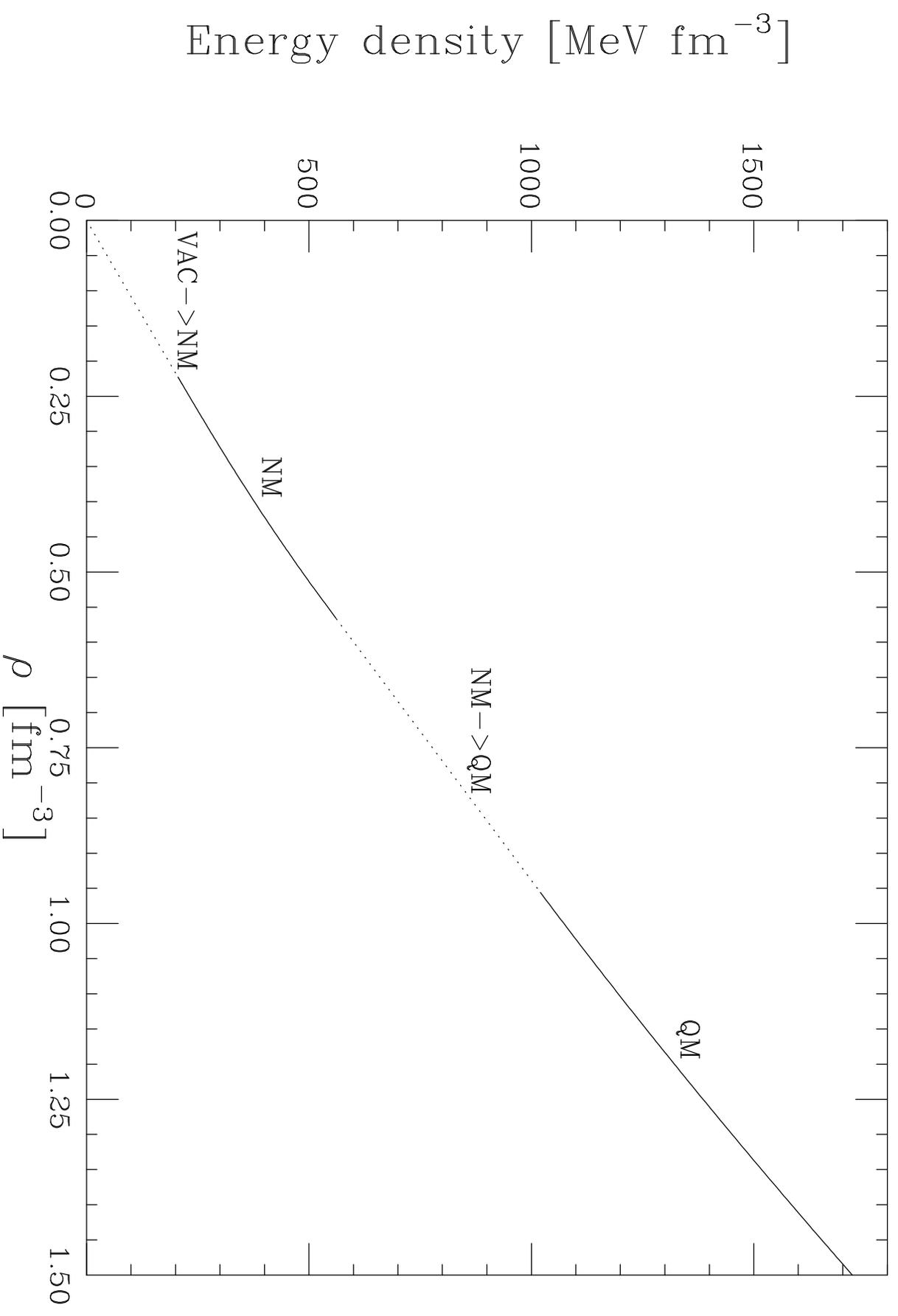
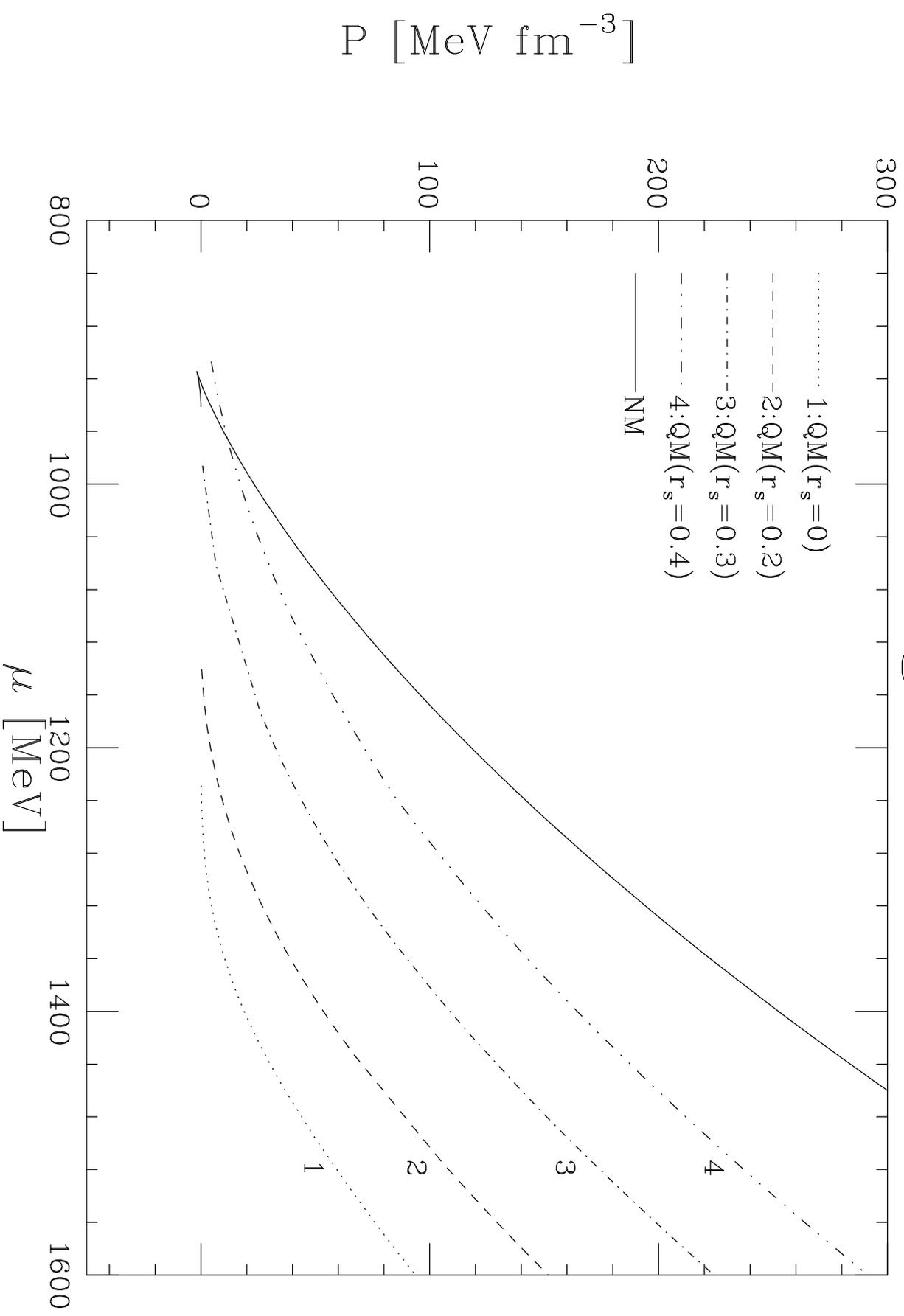
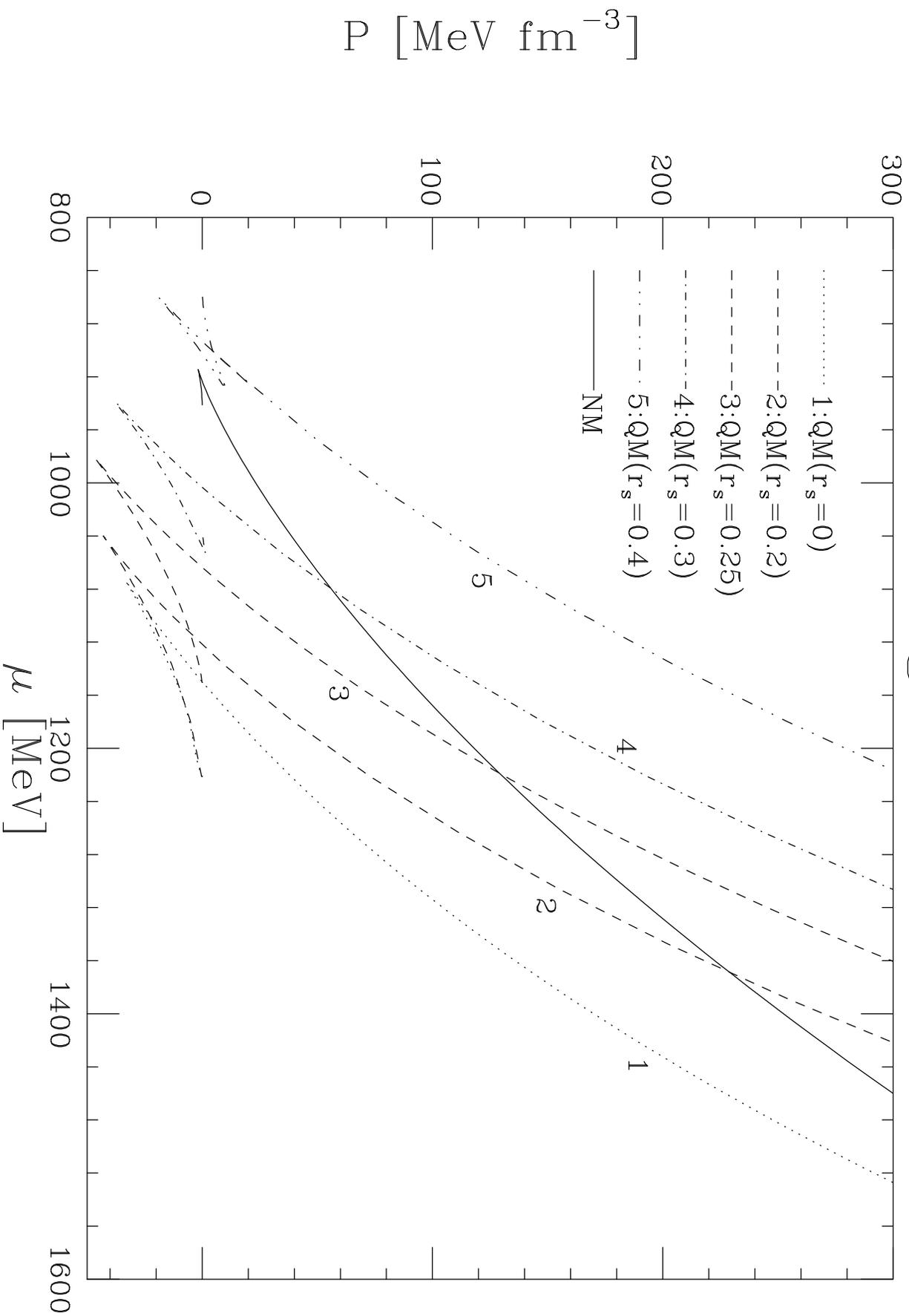
\end{document}